\documentclass[12pt]{article}\usepackage[hyperfootnotes=false]{hyperref}
\usepackage{epsfig}

\usepackage{floatrow}
\usepackage{caption}

\usepackage{amsmath}
\usepackage{amssymb}
\usepackage{graphicx}
\newlength{\abstractwidth}
\renewcommand{\thefootnote}{\fnsymbol{footnote}}
\renewcommand{\thanks}[1]{\footnote{#1}} % Use this for footnotes
\newcommand{\starttext}{
\setcounter{footnote}{0}
\renewcommand{\thefootnote}{\arabic{footnote}}}
\renewcommand{\theequation}{\thesection.\arabic{equation}}
\newcommand{\be}{\begin{equation}}
\newcommand{\bea}{\begin{eqnarray}}
\newcommand{\eea}{\end{eqnarray}}
\newcommand{\beq}{\begin{equation}}
\newcommand{\ee}{\end{equation}}
\newcommand{\eeq}{\end{equation}}

\def\ba{\begin{eqnarray}}
\def\ea{\end{eqnarray}}

\def\12{{1 \over 2}}
\def\eq{&=&}

\def\la{\langle}
\def\ra{\rangle}

\def\simleq{\; \raise0.3ex\hbox{$<$\kern-0.75em
\raise-1.1ex\hbox{$\sim$}}\; }
\def\simgeq{\; \raise0.3ex\hbox{$>$\kern-0.75em
\raise-1.1ex\hbox{$\sim$}}\; }

\def\O2{\Omega_2}

\def\bi{\begin{itemize}}
\def\ei{\end{itemize}}

\def\sc{\setcounter{equation}{0}}

\def\W{$\Omega$}
\def\W'{$\Omega$}

\def\V{\Omega}
\def\V'{\Omega}

\def\O{${\cal{O}}$}

\def\c{{\cal{C}}}

\def\bn{\bigskip \noindent}

\def\suk{SU(2^K)}

    \def\cg{$\c$-geometry}
     \def\cg2{$\c_2$-geometry}

\makeatletter
\g@addto@macro\normalsize{%
  \setlength\abovedisplayskip{10pt}
  \setlength\belowdisplayskip{20pt}
  \setlength\abovedisplayshortskip{10pt}
  \setlength\belowdisplayshortskip{20pt}
}
\makeatother

\usepackage{color}

\begin{document}
\renewcommand{\theequation}{\thesection.\arabic{equation}}
\begin{titlepage}
\rightline{}
\bigskip
\bigskip\bigskip\bigskip\bigskip
\bigskip
\centerline{\Large \bf { Entanglement is not Enough}}

\bn

\bigskip
\begin{center}
\bf  Leonard Susskind  \rm

\bigskip

Stanford Institute for Theoretical Physics and Department of Physics, \\
Stanford University,
Stanford, CA 94305-4060, USA \\
\bigskip

%\vspace{1cm}
\end{center}

\begin{abstract}

This is the written version of a lecture given at KITP in Oct 2014 on Black Holes and quantum complexity.  I've  included (in boldface) various questions that came up during the lecture and  discussions the following day, as well as the quantitative calculations that form the basis of the arguments.

\medskip
\noindent
\end{abstract}

%\begin{figure}[h!]
%\begin{center}
%\includegraphics[scale=.5]{bnw.pdf}
%\caption*{ }
%\label{}
%\end{center}
%\end{figure}

\end{titlepage}

\starttext \baselineskip=17.63pt \setcounter{footnote}{0}
\tableofcontents

\sc
\section{Computational Complexity}\label{CompComp}

Recently, while I was visiting KITP to give a seminar, I was repeatedly asked why I believe something as absurd as ER=EPR. My answer was: ``It was proved by AMPS." I had thought that it was obvious,
 but it wasn't. I didn't get to explain it till the day after the lecture at a group lunch:

\bn

\bf

What AMPS \cite{Almheiri:2012rt} showed was that after the Page-time the modes ($A$)  just behind the horizon can only be entangled with the modes in front of the horizon ($B$) if they are identified with the radiation  ($R$). It's sometimes called $A=R_B.$ But AMPS pointed out that this has a consequence, namely that if Alice does certain actions on the radiation (very far away from the black hole) she can effect the modes just behind the horizon, and create  particles that Bob would encounter when he falls into the black hole\footnote{Harlow and Hayden  \cite{Harlow:2013tf} showed that Alice's experiment is exponentially complex. That paper was the first to use quantum complexity in black hole physics.  }.  But (said AMPS) that is impossible; the radiation is so far away that it can't possibly have an effect near the black hole. Therefore the particles must have been there whether or not Alice acted on the radiation. Since that has to be true for all modes it must mean that there is a firewall. This was all based on the assumption that there is no causal  ``short-cut"  between the radiation and the interior of the black hole. The same argument should hold whenever the black hole is maximally entangled with any distant object---call it the purifier.

Maldacena and I took another position \cite{Maldacena:2013xja}. We said that AMPS had proved that there must be a short-cut  between the purifier and the  interior of the  black hole. Then you can just say that Alice's disturbance of the purifier sent a signal through the short-cut. If she hadn't made the disturbance no particle would have appeared behind the horizon.

What is the short-cut between the interior of the black hole and the system it's entangled with? Answer: the Einstein-Rosen bridge. For that reason, I explained that  the legacy of AMPS will not be firewalls; it will be ER=EPR.

\rm

\bn

ER=EPR says that two systems are connected by an Einstein-Rosen bridge (ERB) if and only if they  are entangled\footnote{This is a special case of the relation between spatial connectivity and entanglement discussed in a seminal paper by Van Raamsdonk \cite{VanRaamsdonk:2010pw}.}.  Entanglement can't be used to send messages a-causally and neither can ERBs because the ERBs are non-traversable. The entanglement entropy is the minimal cross-sectional  area of the ERB. This is an example of the relation between entanglement and the connectivity of space that has been such a popular subject lately. But the story is not complete. The ERB is not a static object. It grows  linearly with time for a very long time. Classically it grows forever. Quantum mechanically the growth is probably limited by Poincare recurrences.

On the other hand the black hole comes to thermal equilibrium quickly. If it's an ADS black hole the dual system comes to equilibrium after a few ADS times and then stops evolving in any obvious way. That leads to the following question: If the evolution stops when thermalization occurs, what is there  in the dual theory  that can represent the continuing growth of the ERB?   This dilemma is closely related to one of the AMPS arguments.

To answer this I will need to explain a few things about  computational complexity.
Computational complexity is a concept from computer science that has to do with quantifying the difficulty of carrying out a task. The ingredients are a system, a space of states, a concept of a simple state, a concept of simple operations, and a task. The task is generally to start with a simple state and transform the system to some other state. Computational complexity  answers the question: What is the minimum number of simple operations  required to carry out the task? I'll illustrate it with a classical example first. The system is a set of $K$ classical bits---call them coins. Each coin can be in one of two states, heads $0$
or tails $1.$ The space of states is described by $K$ binary digits,  $(00110110001....).$ For definiteness let's make an identification under the $Z_2$ operation that flips all the coins so that $(0110001...) = (1001110...).$

The obvious candidate for a simple state is all heads $(00000..)$ ( or equivalently, all tails). By simple operations we mean operations that involve a small number of coins at a time. The simplest operation is to flip a single coin, $(0\leftrightarrow 1).$ Finally the task is to take the system from the simple state to some specific final state.

The complexity of a state is by definition the minimum number of simple operations that are required to carry out the task. It is obvious that no state requires more than $K/2$ flips, so the maximum possible complexity is $K/2.$

Less obvious but true is that almost all states have close to maximal complexity.

The maximal entropy is also proportional to $K,$  namely $K \log{2}.$ Complexity and entropy are not the same thing but they are similar classically.

Complexity is not necessarily the number of steps it takes to do the task by any particular dynamical rule of evolution. It is the least number of steps with any rule that only uses simple operations. For a generic dynamical rule it will take longer, but not that much longer. Typically the time that it takes is the same as the time it takes to get to the maximum entropy, i.e., the thermalization time. The thermalization time is generally short, polynomial bounded in $K.$

A much longer time scale is the Poincare recurrence time. We can think of it as the typical time that it takes to go from a complex state back to the simplest state. It's not the shortest time but the typical time for a generic dynamical rule. Since there are $2^K$ states, the time is of order $2^K.$ To review,

\bea
\rm \it \ \c_{max} \eq K/2  \ \ \ \ \ \ \     \rm  maximum \ complexity \it  \cr  \cr
\rm  \it \   S_{max} \eq K \log{2}           \ \ \ \ \ \ \     \rm  maximum \ entropy \it  \cr  \cr
\rm  \it \  t_{therm} &\sim& t_{comp}    <  K^p    \ \ \ \ \ \ \     \rm  time \ to \ thermalize \ or \ get \ maximally \ complex           \it  \cr  \cr
\rm  \it \   t_{rec}    \eq   e^K   \ \ \ \ \ \ \     \rm   recurrence \ time         \it
\label{classical list}
\eea

\bf

\bn

Question: Do you consider the state $(01010101010101...)$ to be very complex? It only takes a few words to define it, so in some sense it's simple.

\bn

Answer: There are two distinct concepts of complexity. One of them has to do with how much information does it take to specify a task. In your case the task is to bring the system to a state in which the c-bits alternate, from a state in which they are all the same. That took twenty-one English words so it can't be very complex. It also doesn't depend on the number of c-bits.

The other concept is how many operations does the machine have to do to carry out the task. That clearly does depend on $K.$ It's the difference between how long the program must be, and how long it has to run.  It's in the latter sense that $(01010101010101...)$  is complex.

\bn

\rm

Now let's turn to the quantum case. The system is $K$ qubits which at first sight seems not much different from $K$ coins. The states have the form,

\be
|\psi\ra = \sum_1^{2^K} \alpha_i |i\ra
\label{feynman}
\ee
Instead of $K$ binary digits it takes $2^K$ complex numbers to specify a state. That's the origin of the huge difference between classical and quantum complexity  \cite{Feynman:1981tf}.

The quantum version of complexity also involves a concept of a simple state, simple processes, and a task.
Simple states have no entanglement among the qubits. The simplest would be a product state with all qubits in the same state. If we identify states under a global $SU(2)$ rotation then there is a unique simplest state,

\be
|0\ra = |00000....\ra
\ee

Simple processes are a little more complicated than the classical case. First of all by a quantum process I mean acting with a unitary operator. By simple I mean operators that involve a very small number of qubits. However, you can't get very far with one-qubit operators since they will never create entanglement. So we must include two-qubit operators. It turns out that two-qubit operators are enough. Simple unitary one or two qubit operators are called gates. A sequence of gates is called a quantum circuit.

The task is straightforward to state if not to carry out. There are two versions. The first is to produce a given $K$-qubit operator $u$ by a sequence of gates.

\be
u = g_n g_{n-1}.....g_1
\ee
 The complexity is defined as the minimum size of a circuit that can produce $u$ in this way.

 A better version is to let each step involve $K/2$ gates, so that each qubit is involved in a single gate in each time step.
 This is a kind of parallel computing that more closely resembles Hamiltonian evolution. The complexity is still defined to be the number of gates in the minimal circuit.

We can also speak of the complexity of a state. The state $|\psi\ra$ can be obtained from the simplest state by action of a unitary,

\be
|\psi\ra = u |0\ra
\ee
The complexity of $|\psi\ra$ is the minimum complexity of any unitary that does the job.

Because of the exponential number of parameters  quantum states can be much more complex than classical states. I'll write a table similar to \ref{classical list} that illustrates this.
\bea
\rm \it \ \c_{max} \eq e^K  \ \ \ \ \ \ \     \rm  maximum \ complexity \it  \cr  \cr
\rm  \it \   S_{max} \eq K \log{2}           \ \ \ \ \ \ \     \rm  maximum \ entropy \it  \cr  \cr
\rm  \it \  t_{therm}  &<&  K^p                \ \ \ \ \ \ \     \rm time \ to \ thermalize \it  \cr  \cr
t_{comp}    \eq   e^K \ \ \ \ \ \ \     \rm  time \ to  \ get \ maximally \ complex           \it  \cr  \cr
\rm  \it \   t_{rec}    \eq   e^{e^K}   \ \ \ \ \ \ \     \rm   recurrence \ time         \it
\label{quantum list}
\eea
As for the first equation in \ref{quantum list}, one additional point is that almost all states are exponentially complex. The statements about time scales for thermalization, maximum complexity, and recurrences are assuming the actual evolution of the system is generated by what I'll call an easy  Hamiltonian. An easy Hamiltonian is one that is a sum of simple Hermitian operators: a simple Hermitian being one involving a small number of qubits---to be specific one and two-qubit terms. The evolution by easy Hamiltonian is analogous to evolution be a quantum circuit composed of simple gates.

The thing to notice is the spectacular difference between the classical and quantum maximal entropy and maximal complexity. The time $t_{therm}$ is the time to achieve maximal entropy, while the time $t_{comp} $ is the time to achieve maximal complexity. Quantum mechanically those times are vastly different.
  What this proves is that there are subtle changes that take place in a chaotic quantum system long after it has come to thermal equilibrium. Complexity is a real property of a quantum state, but normally we are not interested in it because the information that it encodes does not show up in ordinary local properties. However, it seems that the incredibly subtle correlations encoding complexity correspond to global unsubtle properties of the inside geometry of black holes \cite{Susskind:2014rva}\cite{Stanford:2014jda}\cite{Susskind:2014jwa}\cite{Roberts:2014isa}.

\begin{figure}[h!]
\begin{center}
\includegraphics[scale=.3]{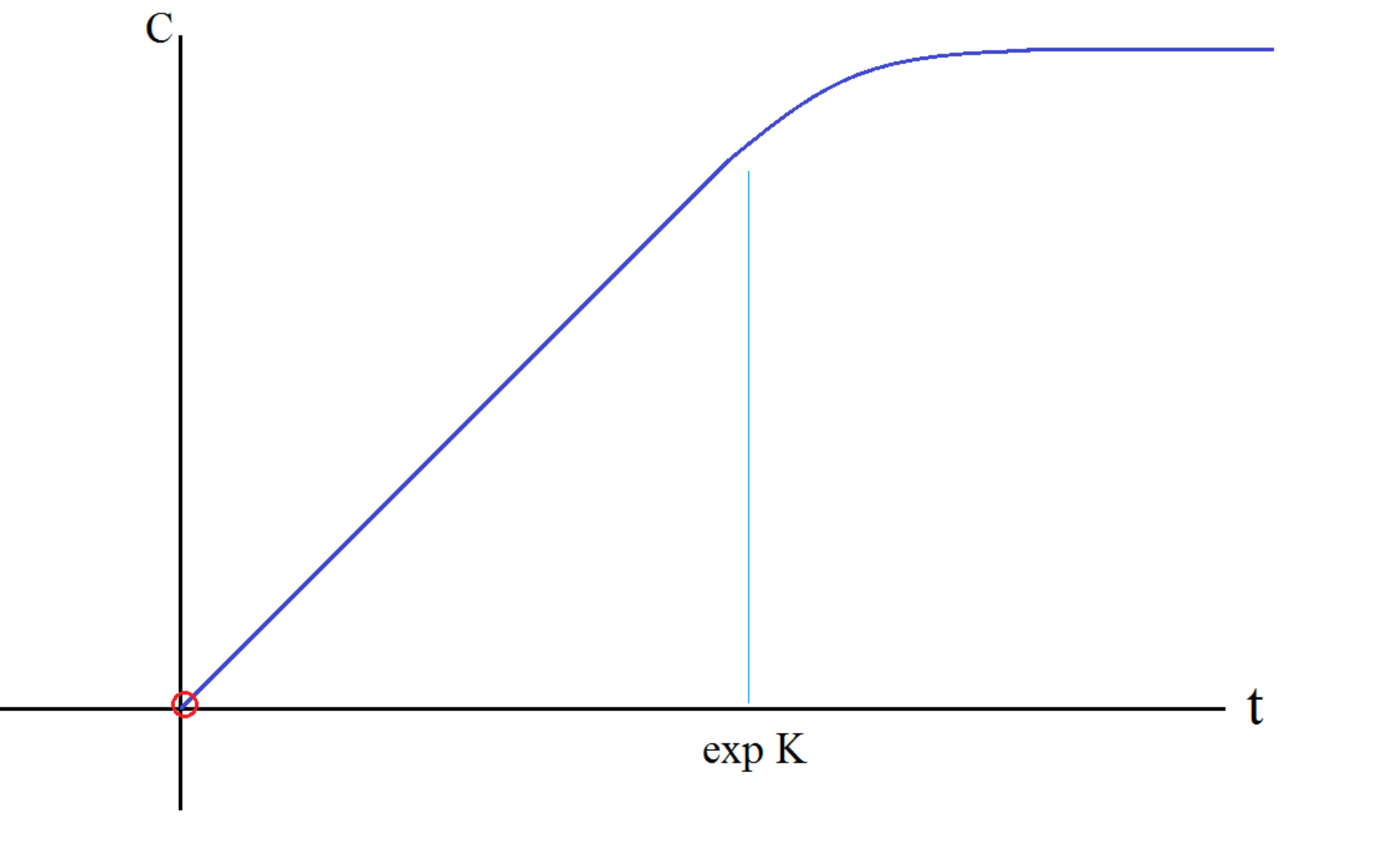}
\caption{ }
\label{A}
\end{center}
\end{figure}

Another point is that the state of a system does not become generic at the thermalization time. It takes an exponential time to reach the complexity of a generic state. A graph of the increase of complexity for a typical chaotic system looks like figure \ref{A}.
It increases linearly for a long time, but since the complexity is bounded by an exponential in $K$ the growth must saturate at
$\log{C} \sim K.$ On the same graph the history leading up to thermal equilibrium would occupy a tiny region shown schematically in the red circle.

It is possible to guess the slope of the curve during the linear growth. The guess is based on two observations. The first is the complexity is extensive; it is proportional to the number of active degrees of freedom of the system. We may use the thermal entropy as a measure of the size of the system. Therefore complexity and its rate of growth should be proportional to the entropy of the system.

The second observation is that the slope is a rate  and should have units of inverse time or energy. One possibility is that the rate of growth is proportional to the total energy of the system, but this can't be right. The ground states of systems such as extremal black holes have both entropy and energy but obviously whatever  complexity they may have is time independent. The right quantity is the product of the entropy and the temperature.

\be
\frac{d\c}{dt} = TS
\label{C-dot = TS}
\ee

\bf
Question: Can \ref{C-dot = TS} be proved or is it a conjecture?

\bn

Answer: A bit of both.  I would refer you to Nielsen's beautiful papers on geometrizing complexity \cite{Nielsen}. Using Nielsen's method one can prove that complexity grows linearly, for some length of time. It is a conjecture that the length of time is as long as possible; namely until the complexity gets near its maximum allowable value \cite{Susskind:2014jwa}.  I think it should not be too hard to show that it is extensive and therefore proportional to $S.$ There are other quantities with units of energy that could replace temperature. The simplest is the energy per degree of freedom. But this would mean that extremal black holes  have a complexity growth proportional to their mass. This doesn't make sense since extremal black holes are ground states and are therefore static. On the other hand the product of entropy and temperature works very well for near extremal black holes.

\rm

\bn

Now let me come to  the punch line.  Earlier I mentioned that ERBs have the property that they eternally grow with time,  at least classically, but all the obvious properties of a chaotic system become static by the time the system thermalizes.
Thermalization times are short, so the question is what property of the quantum state could possibly be dual to growth of the ERB? It can't be entanglement between different parts of the system, because that reaches its maximum   very quickly\footnote{This line of reasoning is related to one of the firewall arguments.}.  As I have explained, thermalization is far from the endpoint  of the evolution of quantum chaotic systems.
The proposal that I have put forward is that the growth of the ERB is dual to the growth of the quantum complexity of the evolving state.

\section{Evolution of ERBs}
\subsection{Nice Slices}

 Classically, any  codimension-one spacelike Cauchy surface     can be divided into two complementary subsets.
The first consists of all points that can send messages to the asymptotic boundary. The second subset is the generalized Einstein-Rosen bridge that lies behind the horizon. The two are separated by a codimension-two surface that marks the intersection of  the generalized horizon with the Cauchy surface.

In order to follow the history of the interior region we will want to foliate the geometry with global spacelike slices. I will impose some restrictions on how the slicing is to be done:

\bi
\item The slices should be Cauchy surfaces. Technically all geodesically complete time-like and light-like curves must intersect each slice exactly once.

    \item The slices should be ``nice slices"  which means they should stay away from singularities and regions of  high curvature.

        \item The slices should be defined, up to a global Lorentz transformation,  in a coordinate invariant way\footnote{The choice of asymptotic Lorentz frame does affect the maximal slices. However the effect dies away at late time as we move deeper into the ERB.}.

        \item At large space-like distances the slices should asymptote to ordinary equal-time surfaces, and be parameterized by a time variable $t$ which coincides with ordinary asymptotic time (as measured by clocks at infinity).

            \item The entire region outside horizons should be foliated by these surfaces. Because they don't enter regions of high curvature they cannot foliate the entire geometry behind the horizon. For our purposes this will not be a disadvantage.

\ei

 Begin by choosing an asymptotic Lorentz frame. At a given time $t$ in that frame, consider the set of space-like slices that are anchored on a spatial sphere at some large value of the radial coordinate $r$. Among these slices there  is one  of maximum spatial volume \cite{Stanford:2014jda}\cite{Susskind:2014jwa}. Letting $r\to \infty,$ this maximal slice defines a unique  global spacelike surface for each value of time. Allowing $t$ to vary foliates the
 the spacetime by such \it maximal slices.\rm   \ In the case of empty space these slices are the usual constant time surfaces in the chosen Lorentz frame.

\begin{figure}[h!]
\begin{center}
\includegraphics[scale=.5]{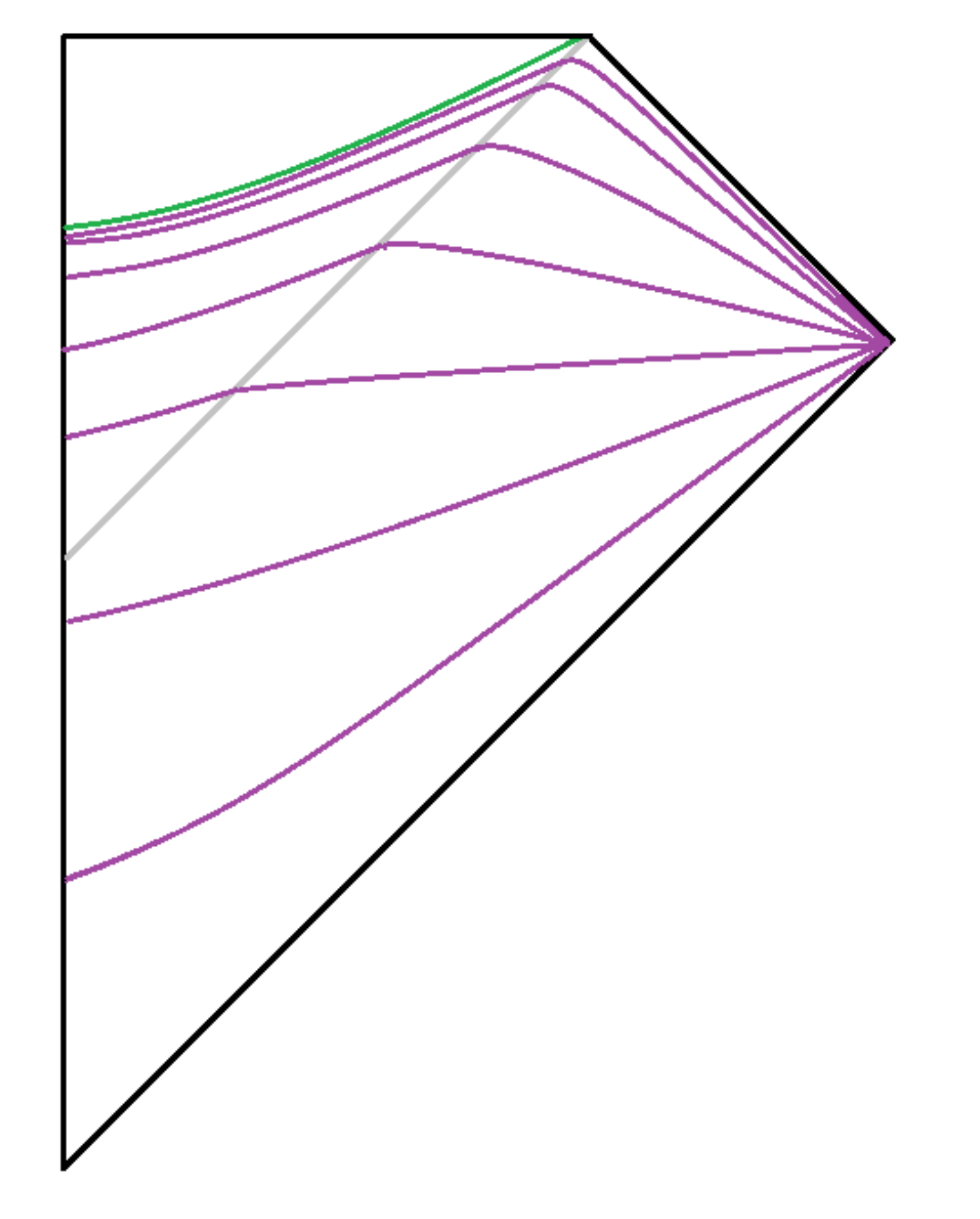}
\caption{Penrose diagram foliated by maximal slices. The green curve is the ``final slice." }
\label{B}
\end{center}
\end{figure}

Now let's introduce collapsing matter to form a black hole; the process is shown in figure \ref{B}

We  follow exactly the same procedure, picking an asymptotic Lorentz frame and anchoring slices on a distant spatial sphere. For each asymptotic time we again choose the slice which maximizes the spatial volume, thus foliating the exterior of the black hole. But inside the horizon the maximal slices pile up before the singularity is reached. The pileup occurs at the \it final slice \rm which is anchored at $t=\infty.$ The final slice is shown as  a green curve in figure \ref{B}. The final slice is independent of the chosen Lorentz frame.

The advantages of maximal slicing are; first, that it is defined in a coordinate invariant manner; and second, the slices are  ``nice slices,"  meaning that all infalling matter passes through them, and they avoid singularities and regions of high curvature.

The disadvantage is that they don't cover the entire spacetime. The foliated region is not geodesically complete because of the pileup on the final slice.  We will not try to evolve the ERB past the final slice.

\begin{figure}[h!]
\begin{center}
\includegraphics[scale=.3]{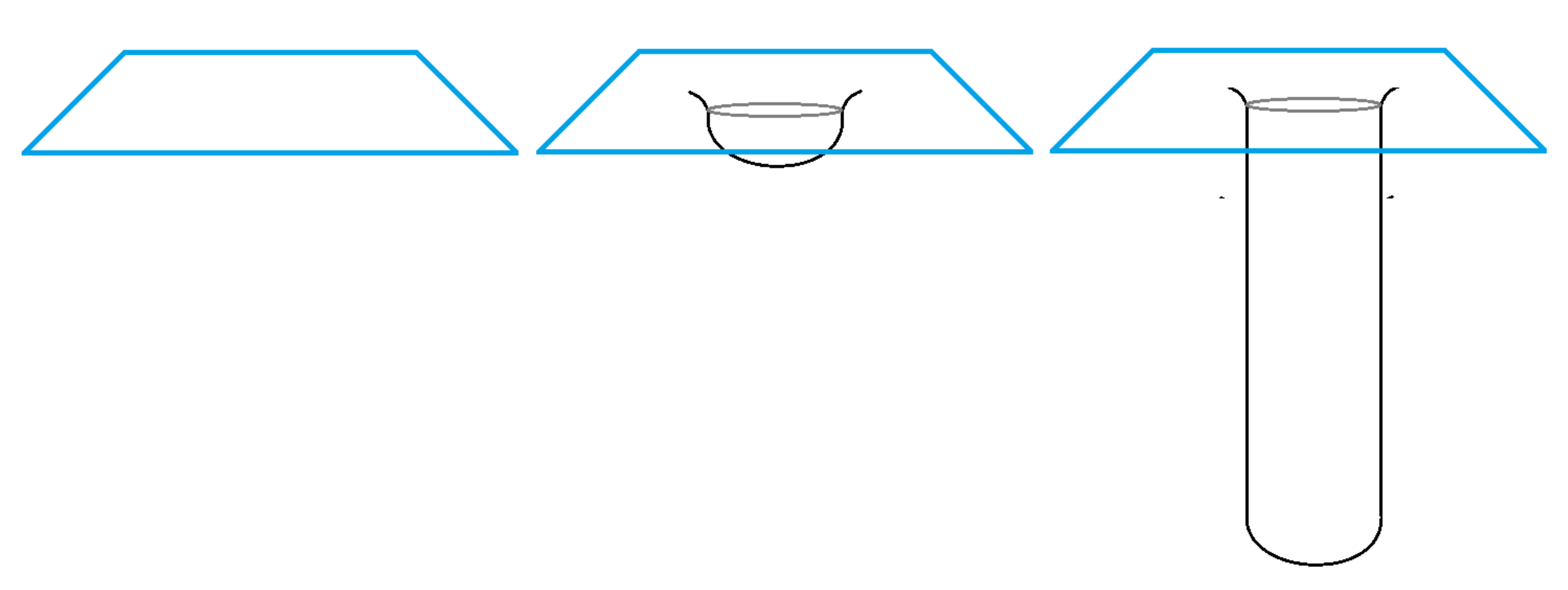}
\caption{ Embedding diagram for an ERB1.}
\label{C}
\end{center}
\end{figure}

Using  the maximal slice at time $t$ we may draw time-dependent embedding diagrams. In figure \ref{C} the  embedding diagrams for a spherically symmetric collapse are shown for an early time and a somewhat later time.

The important features  are these:

\begin{enumerate}

    \item The history begins without  black holes, and the foliation by maximal surfaces starts out close to the foliation of flat or ADS space by constant time slices.  Once the black hole forms the maximal surface penetrates the horizon.

        \item The foliation covers the entire spacetime in the exterior region, but terminates on the final slice inside the horizon.

    \item The spatial geometry of the interior consists of a growing structure---the \it bridge-to-nowhere \rm or \it one-sided Einstein-Rosen bridge \rm (abbreviated \it ERB1 \rm)---which by definition is the portion of the maximal slice inside the horizon.

        \item The far end of the ERB1 contains the matter that formed the black hole. It recedes away from the horizon with a rigid conveyor-belt-like motion. The coordinate length (Schwarzschild coordinate $t$ which behind the horizon is space-like), measured from the far end to the horizon, is equal to the time measured from formation of the black hole.

                \item From the perspective of the horizon all bits of matter stuck to the ERB1 will get  swept away to infinity. A useful slogan is that \it debris is swept away with the tide.\rm

            \end{enumerate}

It is interesting to ask if matter can move against the tide. It can but only temporarily, and not so far that it escapes the black hole. Suppose an infalling particle decays on the slice $t$ and  emits a photon in the backward direction (toward the horizon). The photon will move against the tide for a while, but it will eventually hit the final slice at a point which is infinitely far from the horizon. In other words it is eventually swept back.
Note that the original matter that formed the black hole is concentrated and frozen near the far end of the ERB1.

The equations behind the pictures are straightforward and of two kinds. The first are the Einstein equations for the space-time geometry,

\be
\delta S_{GR}=0
\ee
along with suitable initial conditions.

The second set   are the variational equations and boundary conditions for the maximal slices,

\bea
 \delta V \eq 0.  \cr \cr
V \eq \int d^d \sigma \sqrt{|g_d|}
\label{vary V}
\eea
where $d$ is the dimensionality of space.
The boundary conditions are that the slice is asymptotically anchored at time $t$ in the specified Lorentz of ADS frame.

\begin{figure}[h!]
\begin{center}
\includegraphics[scale=.3]{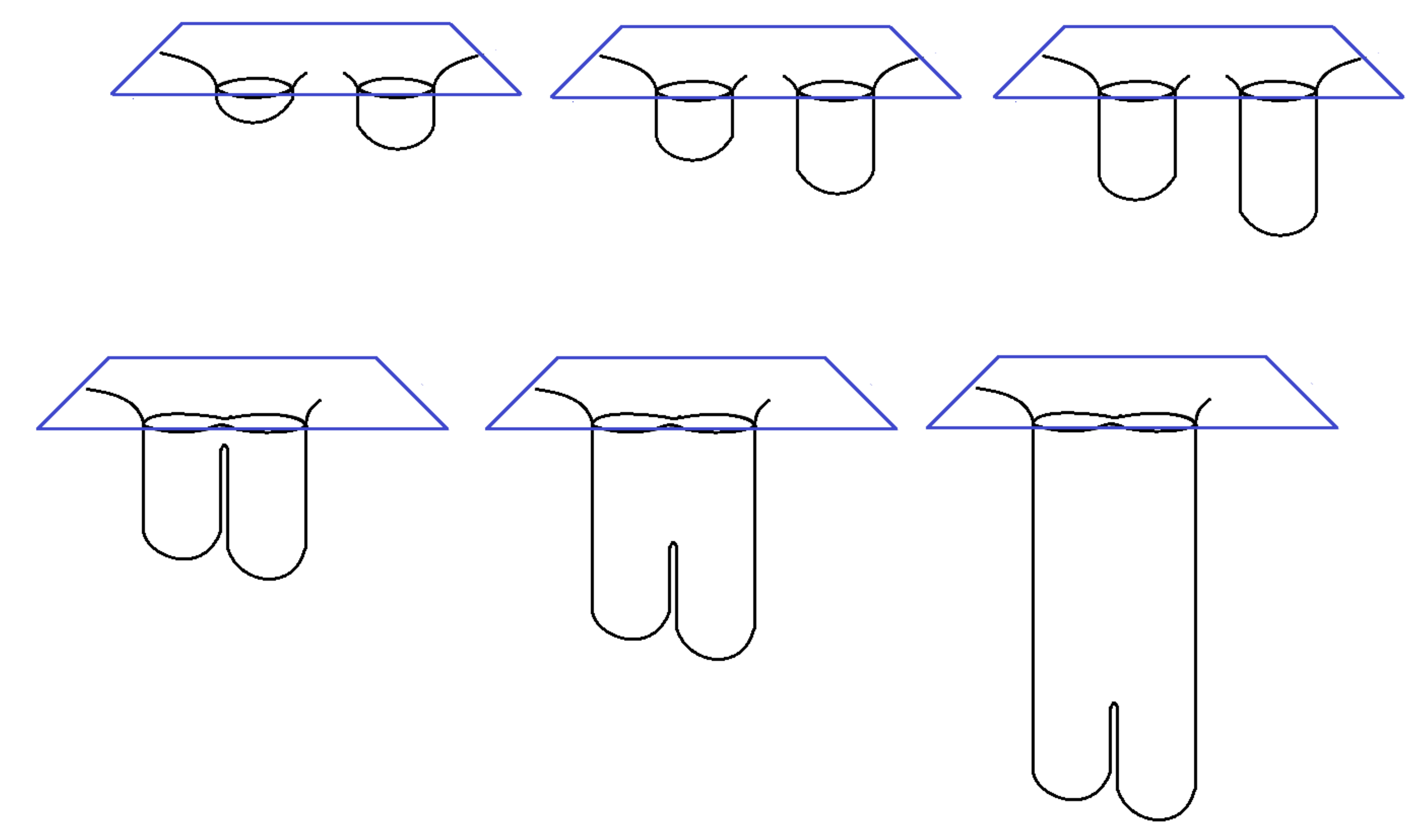}
\caption{ Evolution of the interior of two black holes as the horizons merge. The top row shows the ERBs growing while before the
merger. The bottom row shows the evolution after the merger.}
\label{D}
\end{center}
\end{figure}

It's easy to  consider multiple black holes within the same framework. Several horizon can form and behind each of them and ERB1  will grow. A series of embedding diagrams for two black holes formed at different times are shown in the top half of   figure \ref{D}.
The horizon consists of two disconnected components, each component having the topology of a $(d-1)$-sphere, from which an ERB1 grows.

Black holes may of course fuse---the horizon changing topology---so that a two-hole geometry evolves into a one-hole geometry. Suppose that the two black holes are fairly mature when they collide so that each has a long ERB1. On some slice the two horizons will fuse so that later slices contain a single connected component. At the fusion point the two
ERB1s  join and form an evolving pair of pants with a receding crotch as in the bottom half of figure \ref{D}.

In classical general relativity the topology of spatial slices is conserved with time. This is not completely trivial and  would not be the case if the maximal slices were to intersect the singularity. But this does not appear to happen. The reason is that the $(D-2)$-sphere shrinks to zero which tends to repel the maximal surface.   This is not a proof but I don't know any counterexample\footnote{The construction fails for geometries that contain inflating de Sitter like regions behind the horizon. I thank Don Marolf for a discussion of this point.}.

On the other hand topological transitions are not forbidden by  symmetry principles or conservation laws. Therefore quantum mechanics should permit them.  At the level of macroscopic gravity we may think of topological transitions as localized  tunneling events. There is at least one existence-proof:
the pair-creation of charged near-extremal black holes in a electric field \cite{Garfinkle:1990eq}. The process is mediated by an instanton which is the gravitational generalization of the Schwinger pair-creation instanton. Immediately after nucleation the black holes are at rest and adjacent to one another, separated by about a Schwarzschild radius. After the nucleation event the electric field accelerates the oppositely charged black holes away from each other.

There are two noteworthy features of the black hole pair creation process. The first is that the rate is enhanced relative to the Schwinger formula, by a factor $\exp S$  where $S$ is the entropy of a single black hole. This factor is understood as a sum-over-final-states factor, but the entropy that appears is not the sum of the entropies of the both black holes---just one black hole. The form of the  enhancement factor indicates that the black holes are produced in a maximally entangled state.

The other feature involves the geometry of the instanton and the subsequent continuation to Minkowski signature. When the black holes are produced the two horizons are joined by an Einstein-Rosen bridge. The ERB is similar to the one that  connects the two-sided ADS black holes in the previous section. The instanton induces a transition in the topology of the foliating slices. The ERB then grows in the manner described earlier.
The Einstein-Rosen bridge that connects the entangled pair is an example of the  ER=EPR  principle \cite{Maldacena:2013xja}:

 \bn

 \it

A black hole which is part of an entangled system is connected to its purification by a non-traversable wormhole, or ERB.

\rm

\bn

There are reasons to think this principle is very general. There are undoubtedly many more processes (besides pair creation) in which topology change occurs accompanied by entanglement. Pair creation is the best understood but it possible to make some educated guesses about others. One case is the ``fission" process in which a Schwarzschild black hole splits into two ``daughter" black holes\footnote{The term daughter is used in nuclear physics for the final nuclei in a fission process.}. This is  impossible classically but since there are no symmetries or conservation laws forbidding it, it must be possible quantum mechanically. For example a black hole of  mass $M$ may spontaneously split into two  black holes of mass $M/2.$ If the initial black hole has entropy $S$ then the final black holes will each have entropy $S/4.$ The process involves a large decrease of entropy by amount $-S/2$ and is therefore very rare. Nevertheless it
 can occur  with an exponentially small probability. Of course the black hole will probably evaporate long before it splits, but still, the splitting is a possible mode of decay.

 The following description is guesswork which at present is not supported by calculation. Let's assume that the initial black hole is accompanied by long bridge-to-nowhere before it splits. The splitting process begins with a large thermal fluctuation in which the horizon is distorted into two almost disconnected lobes. The area of the horizon at this point is about half the original area. A reasonable guess for the rate of such fluctuations is $e^{-S/2}.$ (Classically this is zero because the entropy contains an inverse factor of $\hbar.$)

\begin{figure}[h!]
\begin{center}
\includegraphics[scale=.3]{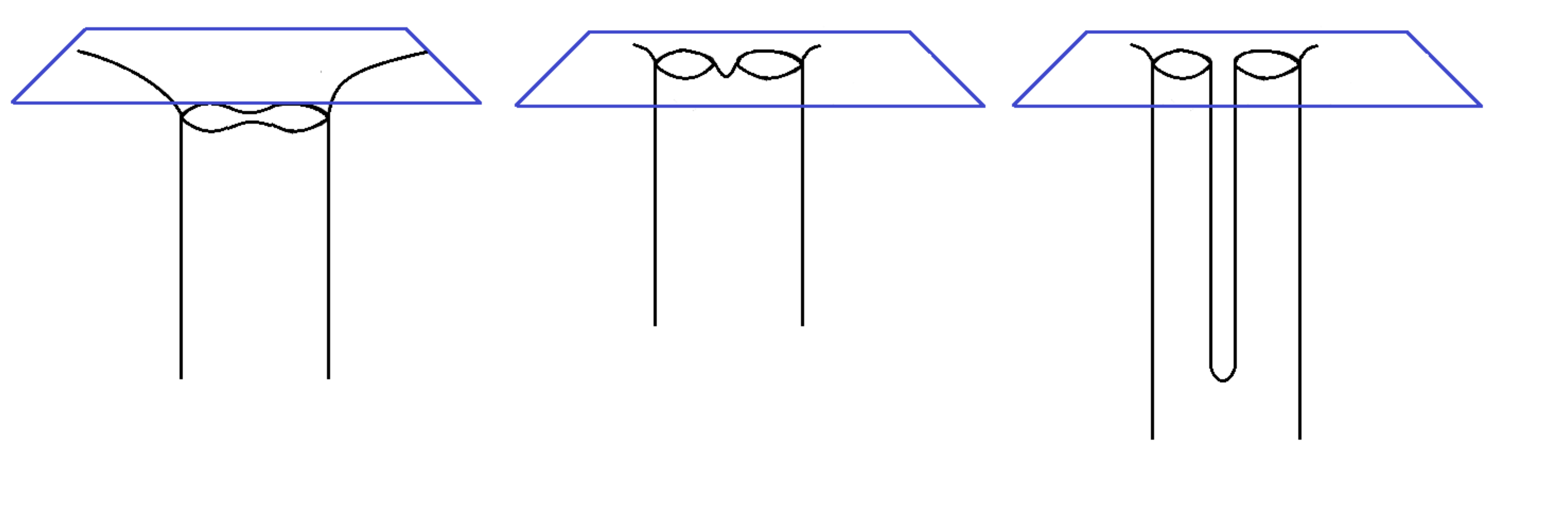}
\caption{Fission of a black hole into two equal daughters. }
\label{E}
\end{center}
\end{figure}

 If we make the usual assumption that the black hole degrees of freedom are thoroughly mixed before the black hole splits then the two daughters will be highly entangled.
From this point a non-perturbative instanton allows  the horizon to split as shown in the
  embedding diagram in figure \ref{E}.  The process then proceeds classically. Each daughter grows its own interior geometry, but the entire configuration remains connected behind the horizon. The bridge-to-nowhere has become a bridge connecting two entangled black holes.

\begin{figure}[h!]
\begin{center}
\includegraphics[scale=.3]{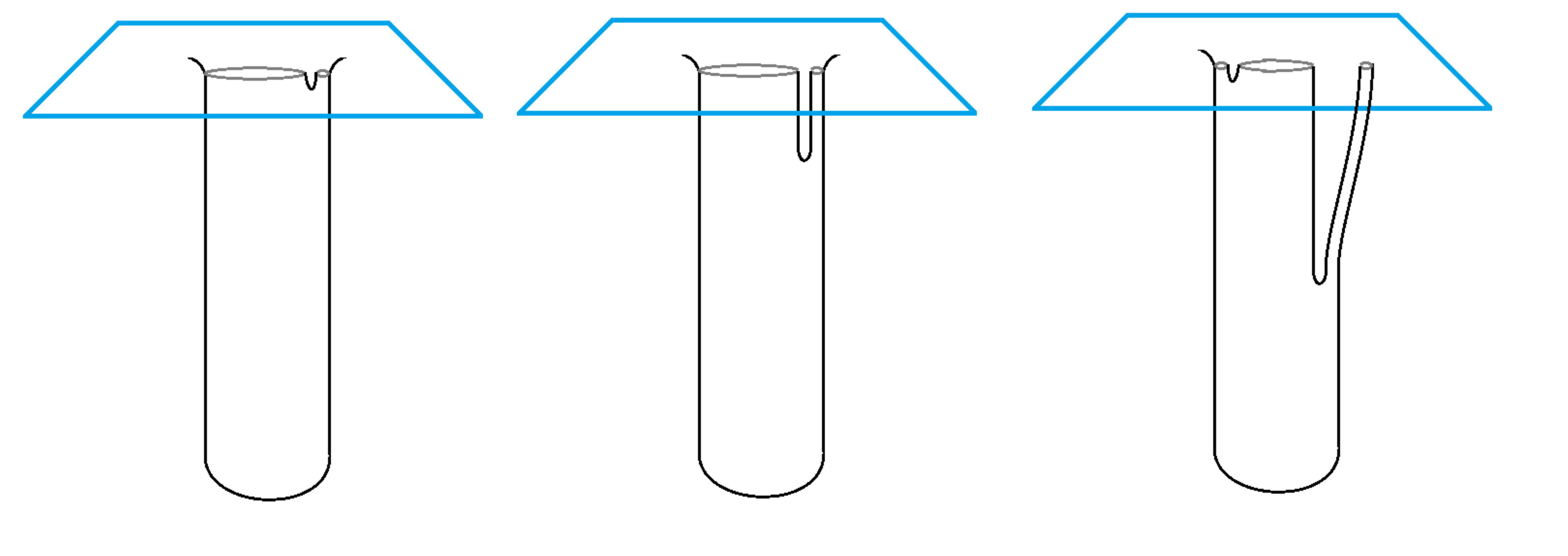}
\caption{Fission into a large black hole and a small daughter. The second panel shows the process repeating. The phenomenon is
the thermal  emission of small black holes in Hawking radiation. }
\label{F}
\end{center}
\end{figure}

For reasons that I will come to it is especially interesting to consider the case in which the black hole splits very asymmetrically into a large component and a very small component. Let the initial mass be $M,$ and the final masses  $m$ and $M-m.$ This asymmetric fission is  much more probable than the symmetric process. The entropy change is $-\Delta S =-\frac{2m}{M}S$ and the series of embedding diagrams for it is shown in figure 7.
The rate for the process is proportional to,

\be
e^{-\Delta S}= e^{-\frac{2m}{M}S}.
\ee
This is easily seen to be the same as

\be
e^{-\Delta S}= e^{-m/T_H}.
\ee
where $T_H$ is the Hawking temperature. The process, described by figure  \ref{F}, is precisely the thermal emission of a small black hole in Hawking evaporation. Consistency of thermodynamics implies that it be possible.

Again, if one assumes that the initial black hole was well scrambled, then any small subset of the degrees of freedom will be
maximally entangled with the rest \cite{Page:1993wv}. Therefore we expect that when very asymmetric fission takes place, the emitted small black hole will be highly entangled with the remaining large black hole. It is  also connected to it by an ERB as indicated in figure  \ref{F}.  The successive emission of multiple small black holes, also shown in figure  \ref{F}  is  possible.

Asymmetric fission has a limit in which the small black hole is replaced by an elementary particle; for example a Hawking photon.
If the ER=EPR principle is general then it means that even in this situation one should view the evaporation products as being connected by some kind of limiting ERB to the parent black hole. Because of its shape I've called it an octopus: the head being the black hole,  tentacles the evaporation products.

\subsection{The Two-sided Case}\label{S two side}

``Interesting" Einstein-Rosen bridges---the ones with non-trivial topologically---are quantum phenomena. When they occur the space-like slice containing them will be multiply connected, with non-contractible loops threading the ERB. But
topology is conserved classically; therefore  topologically complex ERBs  cannot  evolve classically from flat-space  conditions.

Nevertheless we can study interesting ERBs without explicitly introducing quantum mechanics  if we allow multiple asymptotic boundaries. In that case the wormhole-like topology is present for all times. Two examples are the Kruskal extension of the Schwarzschild geometry and the two-sided ADS  black hole \cite{Maldacena:2001kr}\cite{Hartman:2013qma}. Because of their greater symmetry the two-sided cases are easier to analyze than one-sided black holes created by collapse.

\begin{figure}[h!]
\begin{center}
\includegraphics[scale=.5]{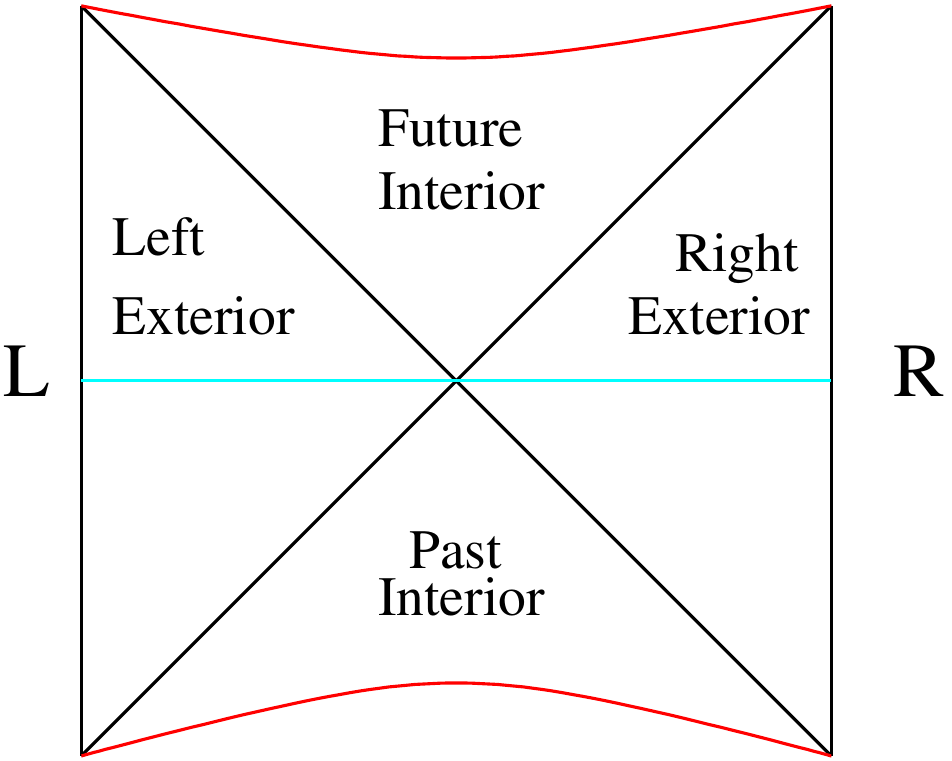}
\caption{Penrose diagram for an ADS two-sided black hole. }
\label{4}
\end{center}
\end{figure}

 In two-sided cases the geometries have two asymptotic regions, each containing an identical black hole. We will refer to the two sides as \it left \rm and \it right. \rm The asymptotic regions  are connected by an ERB.  The non-trivial topology is present from the beginning; thus no transition is necessary.
We will mainly concentrate on  ADS: the
Penrose diagram  is shown in figure \ref{4}.

The metric describing the geometry of  figure \ref{4}
 has the form,

\be
ds^2 = -f(r) dt^2 + f(r)^{-1} dr^2 +r^2 d\Omega_{D-2}^2.
\label{general metric}
\ee
For the  BTZ $3$-dimensional case the function $f(r)$ is given by,

\be
f(r) = \frac{1}{l^2}(r^2 - \mu^2)
\label{f-btz}
\ee
with $l$ being the ADS length scale and
$\mu$ being related to the mass by,

\be
\mu^2 = 8G_N Ml^2.
\ee

For $D>3$ $f(r)$ has the more complicated form,

\be
f(r) = {\frac{r^2}{l^2}} +1 -\frac{\mu^2}{r^{D-3}}.
\label{f D>3}
\ee
with

\be
\mu^2 = \frac{G_NM l^2}{\omega_{D-2}}
\label{mu}
\ee
and $\omega_{D-2}$ being the volume of a unit $(D-2)$-sphere.

\begin{figure}[h!]
\begin{center}
\includegraphics[scale=.3]{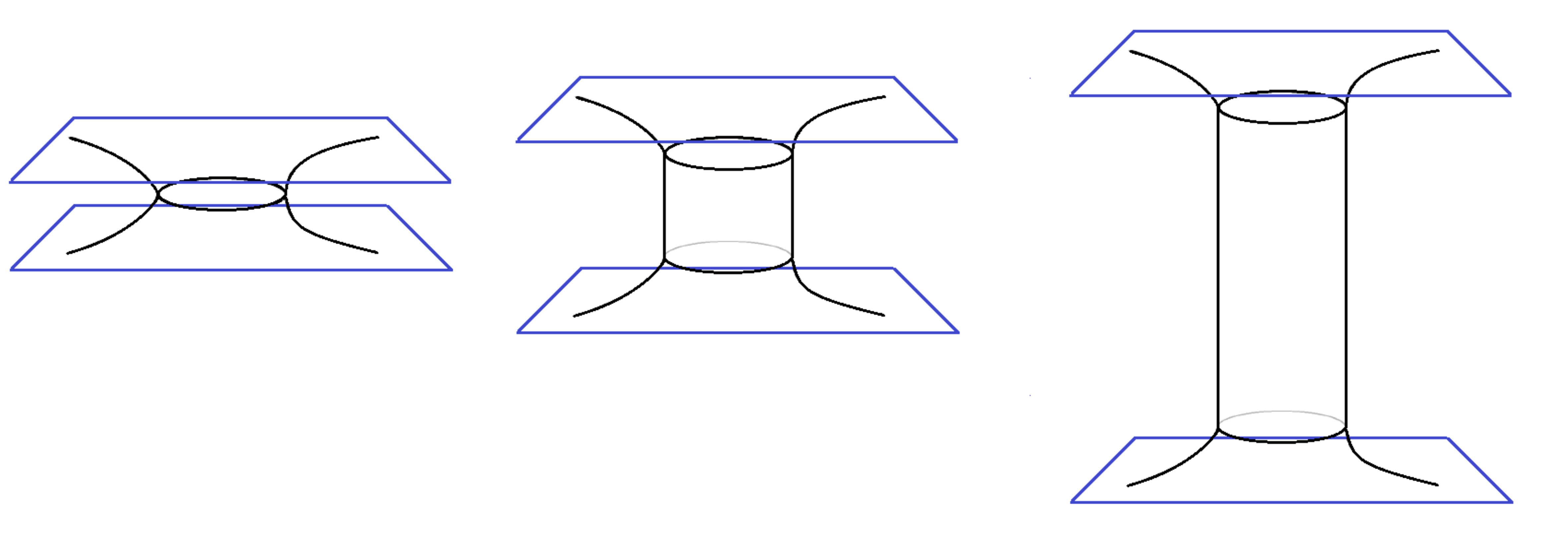}
\caption{ Embedding diagram showing the growth of the two-sided ERB.}
\label{G}
\end{center}
\end{figure}

The full spacetime geometry described by \ref{general metric}
is pathological---it contains white holes as well as black holes on both sides---and is very unstable with respect to perturbations in the lower half of the diagram. For our purposes we can throw away the lower half and assume that the system was created at $t=0$ by some unspecified quantum event, for example a Hartle-Hawking instanton. The history then starts at $t=0$ with the embedding diagram shown in left-most part of figure \ref{G}
As we move up the boundaries of the Penrose diagram, the time  $t$ increases. The embedding diagrams shown in the other two parts of figure \ref{G}  indicate how the ERB grows.

Consider a rotationally invariant slice defined by giving $r$ as a function of $t.$ The volume of such a slice is given by,

\bea
V\eq \omega_{D-2} \int  dt \sqrt{|f(r)|r^{D-2}} \cr \cr
\eq \int dt \sqrt{g(t)}  \cr \cr
g(r) &\equiv&  |f(r)|r^{2(D-2)}
\label{volume formula}
\eea
The equations for the maximal slices are just the variational equations \ref{vary V} applied to the function $r(t).$
They are particularly simple
in the limit $t=\infty.$ In that case  the time translation and rotational symmetry  of the problem imply that the final slice is located  at constant $r.$  The value of $r$ on the final slice is   found  by maximizing the function $g(r)=|f(r)|r^{2(D-2)},$

\be
\frac{\partial g}{\partial r} = 0  \ \ \ \ \ \ (r=r_f)
\ee
Looking at the form of $g(r)$ we see that it is positive and that it vanishes both at the horizon and at $r=0.$ The vanishing at the horizon is due to the usual zero in $f(r)$ that defines the horizon: the vanishing at $r=0$ is due to the fact that the $(D-2)$-sphere shrinks to zero at the singularity faster than $f$ grows. This latter effect is a kind of repulsion which keeps the maximal slices from hitting the singularity. It's interesting to note that this protection does not exist for calculating geodesics crossing from left to right. In general geodesics can crash into the singularity.

In the BTZ case one finds,

 \be
 r_{f} = \frac{\mu}{\sqrt{2}}
 \label{rmax for btz}
 \ee

 The maximal slices for finite $t$ are more difficult to compute. They have spherical symmetry but not $t$-translation invariance. They may be represented by a function $r(s)$ satisfying the variational equation,

 \be
 \delta \int ds \sqrt{|f(r(s))|} \ r(s)^{D-2} = 0
 \label{variational for Vmax}
 \ee
with the boundary condition

\be
\lim_{s\to t}{r(s)} = \infty.
\label{boundary condition r to infinity}
\ee

\begin{figure}[h!]
\begin{center}
\includegraphics[scale=.5]{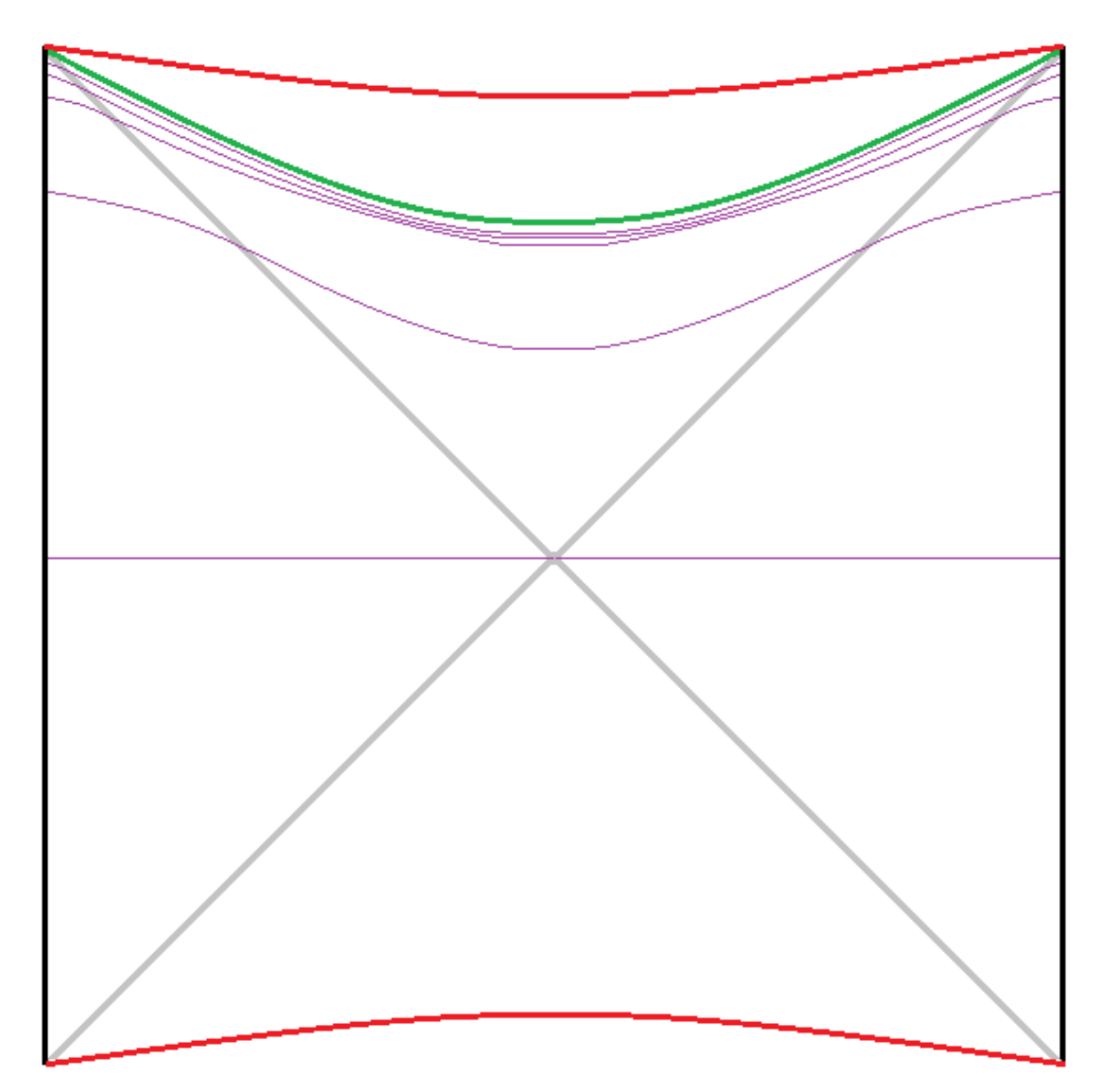}
\caption{Two-sided ADS black hole foliated by maximal slices.  }
\label{H}
\end{center}
\end{figure}
Figure \ref{H} shows the Penrose diagram for BTZ foliated by maximal surfaces.
As the time at which the surfaces are anchored increases the maximal surface moves toward the final slice shown as the green curve. The proper length of the ERB can be defined to be the proper distance, between the left and right horizons, measured along the purple curves. It grows linearly with $t,$

 \be
 \rm length \it \to 2t \sqrt{|f(r_{f})|}
 \label{growth of length}
 \ee
At late time almost all of the ERB volume is very close to the final slice. Only the portion near the ends deviates from $r_{f}.$  Over most of the length the cross-sectional area of the ERB is constant,

\be
A_{max} = \omega_{D-2} r_{f}^{D-2}.
\label{area}
\ee
Combining \ref{growth of length} and \ref{area},  the volume of the ERB grows according to

\be
V = 2 \omega_{D-2} \ r_{f}^{D-2} \  \sqrt{|f(r_{f})|} \ t.
\ee
The area \ref{area} is not exactly the area of the horizon, $A_H$. At the point where the ERB enters the black hole its area---call it $A_H$---is the usual horizon area, but it generally decreases somewhat before becoming constant deeper into the ERB. The geometric  reason is that $r_{f}$ is somewhat less than the horizon radius $r_H.$

\subsection{Additional Formulas for ADS and Schwarzschild}\label{S ads forms}

\bf I will present some quantitative formulas  for the final slice in the ADS and Schwarzschild geometries in space-time dimension $\geq 3.$ They were not covered in my lecture but were the basis for much of it.

First we consider two-sided ADS.
The formulas for  $D>3$ are a bit complicated but if the horizon radius is somewhat  bigger than the ADS radius the constant term in $f(r)$ can be ignored. In that case the formulas have a universal form valid for all $D \geq 3.$ The function $g(r)$ in \ref{volume formula} takes the form

\bea
g(r) \eq \frac{1}{l^2}(\mu^2 Z -Z^2)  \cr  \cr
Z &\equiv& r^{D-1}
\eea
The values of $r_H$ and $r_f$ are determined  by the equations $g=0$ and $g'=0,$ giving,

\bea
Z_f \eq \mu^2/2 \cr \cr
Z_H \eq \mu^2
\eea
 and

 \bea
 r_f \eq \left( \frac{\mu^2 }{2}  \right)^{1/(D-1)} \cr \cr
  r_H \eq \left( \mu^2 \right)^{1/(D-1)}
 \eea

 The function $\sqrt{g(r_f)}$ that appears in the volume formula \ref{volume formula} has the value

 \be
\sqrt{g(r_f)} = \frac{1}{l}|\mu^2 Z_f - Z_f^2|^{1/2} = \frac{\mu^2}{2l},
\label{sqrt g}
 \ee

 The values of the area at the horizon and the  final surface are

 \be
 A_H=\omega_{D-2} \mu^{2(D-2)/(D-1)}
 \label{AH}
 \ee

 \be
 A_f =\omega_{D-2}\left(    \frac{\mu^2}{2}     \right)^{(D-2)/(D-1)}   = A_H \left(\frac{1}{2}\right)^{(D-2)/(D-1)}
 \label{Af}
 \ee
Note that the $D$ dependence in the ratio $A_f/A_H$ is very mild over the entire range of $D.$

The surface gravity of the black hole is given by

\be
\kappa =\frac{f'(r_H)}{2}
\label{kappa}
\ee
where $f' = df/dr.$

One finds

\be
\kappa =\frac{D-1}{2l^2}  \mu^{\frac{2}{D-1}}.
\ee

%%%%%%%%%%%%%%%%%%%%%%%%%%%%%%%%%%%%%%%%%%%%%%%%%%%%%%%%%%%%%%%%%%%%%%%%%%%%%%

%%%%%%%%%%%%%%%%%%%%%%%%%%%%%%%%%%%%%%%%%%%%%%%%%%%%%%%%%%%%%%%%%%%%%%%%%%%%%%

%%%%%%%%%%%%%%%%%%%%%%%%%%%%%%%%%%%%%%%%%%%%%%%%%%%%%%%%%%%%%%%%%%%%%%%%%%%%%%

Let's now consider two-sided Schwarzschild black holes  in a flat background.
The functions $f$ and $g$ are given by,

\be
f(r)=1-\frac{\mu}{r^{D-3}},
\ee

\be
g(r) = \mu r^{D-1}-r^{2D-4}
\ee

The horizon radial coordinate---also the Schwarzschild radius---is

\be
r_H =\mu^{\frac{1}{D-3}}
\ee
and the value of $r$ on the final slice, obtained by maximizing $g,$ is

\be
r_f = \left(    \frac{1}{2}   \frac{D-1}{D-2}    \right)^{\frac{1}{D-3}} \mu^{\frac{1}{D-3}}
\ee

The ration $A_f/A_H$ is again a very mild function of $D,$

\be
\frac{A_f}{A_H} = \left[ \frac{1}{2}    \frac{D-1}{D-2}                 \right]^\frac{D-2}{D-3}
\ee

The surface gravity $\kappa$ is

\be
\kappa =\frac{1}{2} f'_H = \frac{D-3}{2r_H} = \frac{D-3}{2 \mu^{\frac{1}{D-3}}}.
\ee
%%%%%%%%%%%%%%%%%%%%%%%%%%%%%%%%%%%%%%%%%%%%%%%%%%%%%%%%%%%%%%%%%%%%%%%%

\rm

\sc
\section{Growth of Volume and Complexity}
\subsection{Volume-Complexity Relation}
Now I want to turn to the relation between complexity and the volume of the ERB. I will work with the ADS case and just quote results for Schwarzschild.

We can compare complexity and volume  by considering the rate of growth of each. First the complexity: earlier I argued that the rate of growth of complexity (before it saturates) is proportional to the product of temperature and entropy \ref{C-dot = TS}. We can replace this by a geometric quantity,

\be
\dot{\c} \sim \frac{\kappa A_H}{G}
\ee
With that in mind let's consider the rate of growth of another quantity,
 the volume of the ERB. Earlier we saw that for ADS the volume grows according to,

\be
dV/dt = \omega_{D-2}\sqrt{g(r_f)}.
\ee

Using \ref{sqrt g},

\be
dV/dt = \omega_{D-2} \frac{\mu^2}{2l}
\ee
Using \ref{AH} and \ref{kappa} one finds,

\be
dV/dt = \frac{l}{D-1}   A_H \kappa
\ee

Now recall that the surface gravity $\kappa$ and the horizon area $A_H$ are related to the temperature and entropy by,

\bea
\kappa \eq 2\pi T  \cr \cr
A_H \eq 4G_N S.
\eea
Accordingly,

\be
dV/dt =  \frac{8\pi G l}{D-1}  ST
\label{vdot}
\ee

Ignoring numerical factors the  relation between  volume and complexity is:

\be
\c \sim   (D-1)  \frac{V}{ G_N l} .
\label{c-v relation}
\ee
Later we will see that the factor $D-1$ has an interesting meaning.

The relation for the Schwarzschild black hole is similar. If we identify the rate of complexity growth with $TS$
we find,

\be
\c \sim \frac{(D-3) V}{G r_H}  \left[ \frac{1}{2^{\frac{1}{(D-3)}}} \sqrt{\frac{D-3}{D-1}}  \left(   \frac{D-1}{D-2}   \right)^{\frac{D-2}{D-3}}            \right].
\label{c-v relation-sch}
\ee

The  $D$-dependence in the square bracket looks complicated but it is extremely  mild; over the entire range $4\leq D <\infty$  it varies by a factor of $$ \frac{9\sqrt{3}}{24} = .65 .$$ If we ignore it the formula takes the simple form,

\be
\c \sim \frac{(D-3) V}{G r_H}  .
\label{c-v relation-sch-simple}
\ee

As in \ref{c-v relation} there is a linear $D$ dependence in \ref{c-v relation-sch-simple}.

The formula \ref{c-v relation-sch} is similar to  \ref{c-v relation} except that the factor of inverse ADS scale $1/l$ is replaced by the inverse Schwarzschild radius $1/r_H.$

This demonstrates an important fact: The proportionality between complexity and volume is \it not \rm universal. For example, suppose  two Schwarzschild black holes of different mass are present. The overall complexity is not proportional to $V_1+V_2$ but rather to

$$\frac{V_1}{\mu_1} + \frac{V_2}{\mu_2}.$$

\bn

\subsection{Causal Patches}
To find a universal geometric interpretation of complexity we need to understand the ERB as a collection of causal patches.
A causal patch is the causal past of a point on the future boundary of spacetime. For a one-sided black hole all points outside the horizon are in the causal past of time-like infinity. Therefore the entire exterior region is a single causal patch. For the two-sided case there are two exterior regions, defining two causal patches. ERBs are much richer in causal patches, resembling in some ways eternally inflating space-times.

Inside the black hole the future boundary is the singularity.  I will define a causal patch  as follows:
Pick a value of $t$ on the singularity and consider the infinitely small $(D-2)$-sphere at that point. The causal patch consists of the union of the interiors of all past light cones from points in that sphere.

We will consider the intersection of the causal patch and the final slice. This defines the portion of the ERB contained within the causal patch. I will abbreviate it by $CP(t)$---the causal patch of the point $t$---keeping in mind that it refers to the \it  intersection of the causal patch  with the final slice.\rm  \ The topology  of $CP(t)$ is the product of a $(D-2)$-sphere and a finite line interval. For example in $(2+1)$-dimensions it is a finite piece of an ordinary cylinder. The $(D-2)$-dimensional area of the base of the cylinder is $A_f.$ The coordinate length  of the cylinder $\Delta t$ is,

\be
\Delta t = 2\int_0^{r_f} \frac{dr}{|f(r)|}
\label{CP lenght}
\ee

For the $(2+1)$-dimensional case.

\bea
\Delta t \eq \frac{l^2}{2} \int_0^{\mu/\sqrt{2}} \frac{dr}{\mu^2-r^2} \cr \cr
\eq \frac{l^2}{2\mu} \log{ \left(   \frac{\sqrt{2} +1}{\sqrt{2} -1}         \right)     }
\label{CP lenght D}
\eea

The proper length of $CP(t)$ is given by $\sqrt{|f(r_f)|} \Delta t ,$

\be
L = \sqrt{|f(r_f)|} \  \Delta t = \frac{l}{\sqrt{2}} \log{ \left(   \frac{\sqrt{2} +1}{\sqrt{2} -1}         \right)     }
\label{propleng}
\ee

Notice that in \ref{propleng} the mass parameter  $\mu$ drops out completely, and that the proper length of $CP(t)$ is proportional to the ADS radius $l$

More generally in $D$ dimensions the proper length of $CP(t)$ is

\be
L = \dfrac{2l}{D-1} \int_0^{1/2} dy    \dfrac{ y^{{(D-2})/({D-1})} }{y-y^2}
\label{propD}
\ee
The dependence of the numerical integral on $D$ is very weak, and for large $D$ the length $L$ scales like

\be
L\sim \frac{l}{D-1}.
\label{propDapprox}
\ee
The coordinate length scales like

\be
\Delta t\sim \frac{l}{(D-1)\sqrt{|f(r_f)|}}.
\label{coordD}
\ee

The situation for Schwarzschild is similar with $l$  replaced by the Schwarzschild radius $r_H.$

\subsection{Unspooling Complexity}\label{S unspool}

\begin{figure}[h!]
\begin{center}
\includegraphics[scale=.3]{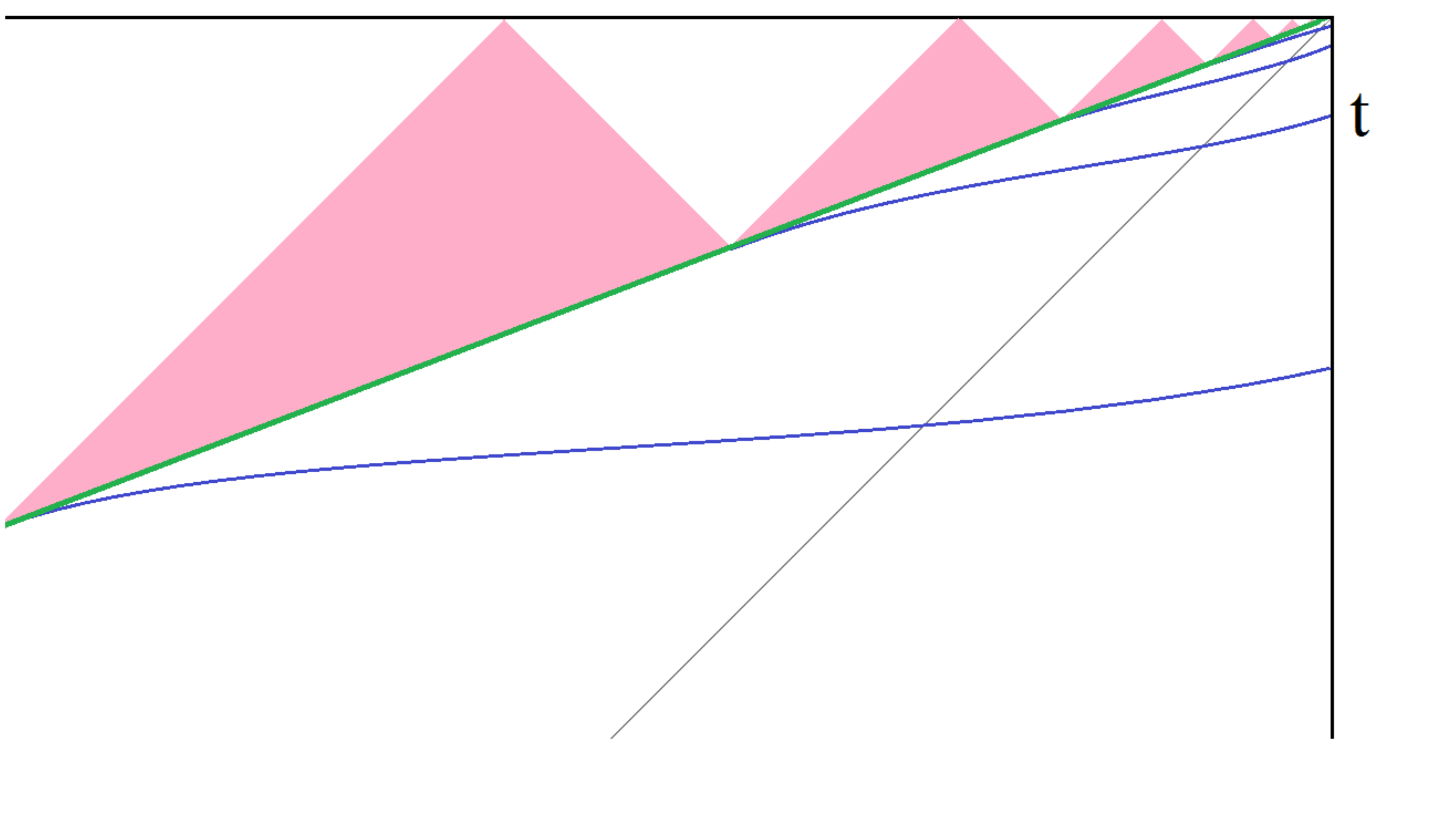}
\caption{The causal patches of points on the singularity intersect the maximal slices and form a chain of $CP(t)$s that unspool from the horizon. }
\label{I}
\end{center}
\end{figure}

\begin{figure}[h!]
\begin{center}
\includegraphics[scale=.3]{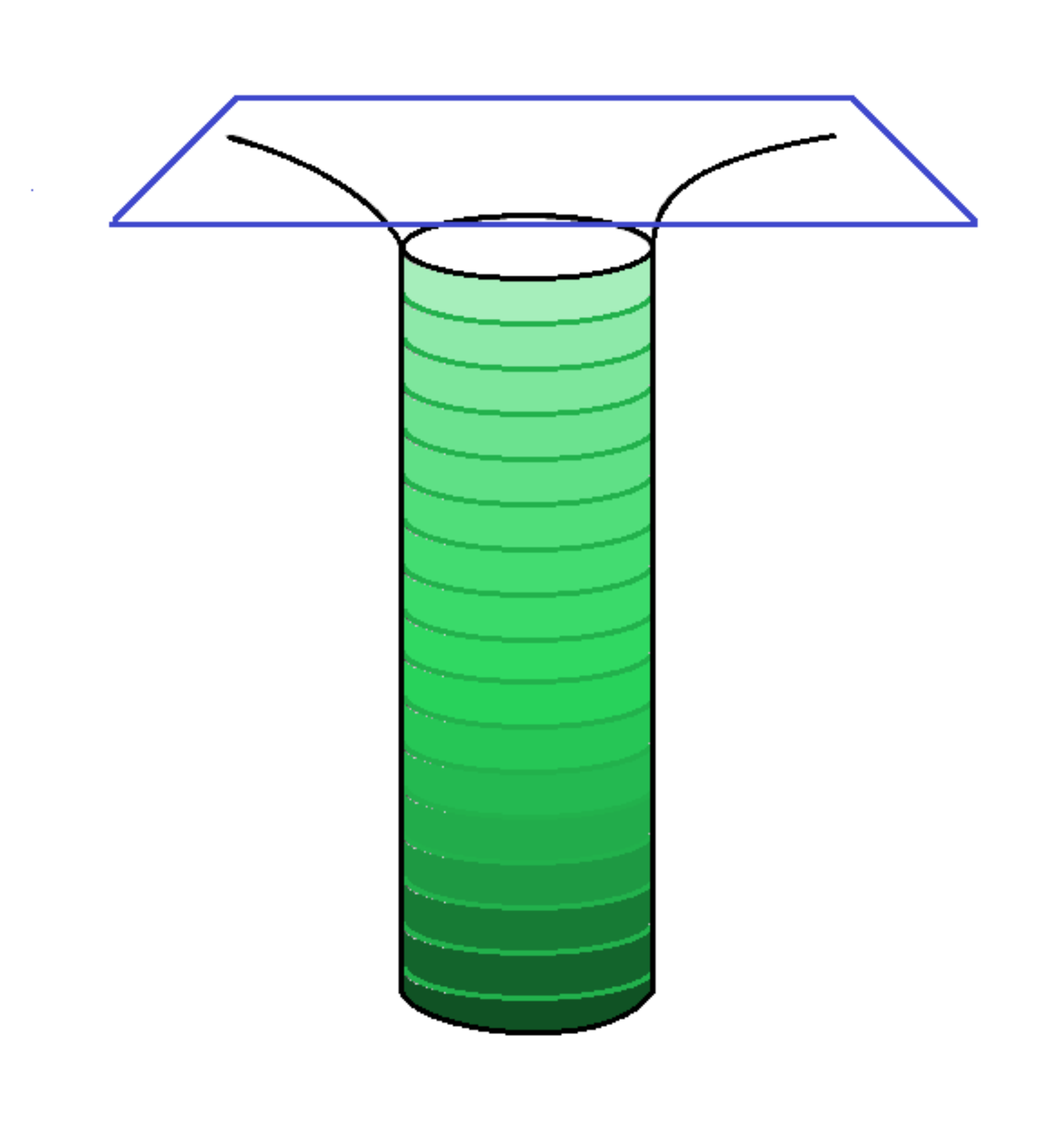}
\caption{An embedding diagram showing the unspooling of causal patches. The darkest patches were the earliest to unspool. }
\label{J}
\end{center}
\end{figure}

For  late but finite $t$ the maximal surface hugs the final slice until it departs near the horizon. The total length of the ERB (The portion behind the horizon) is finite, but grows with $t.$ The creation of new space takes place at the horizon with a motion that I will call ``unspooling." The unspooling can be seen in figure \ref{I}.
It is clear that the unspooling (which takes place in the upper righthand corner) is accompanied by the creation of new causal patches which recede away from the horizon as they are produced. This is shown as an embedding diagram in figure \ref{J}. The dark (light) green bands represent the earliest (latest) causal patches.
The rate of production of  causal patches, is one per $\Delta t.$ Thus after time $t$ the number of non-overlapping causal patches that have unspooled is of order

 $$t\mu/l^2.$$

The final property of a causal patch that we will need is its entropy, or strictly speaking the  entropy bound for the causal patch \cite{Fischler:1998st}\cite{Bousso:1999xy}. One sees from figure \ref{I}  that each causal patch is bounded by two shrinking light sheets which originate on $(D-2)$-spheres of area $A_f.$  The covariant entropy   bound \cite{Bousso:1999xy} on $CP(t)$ is the sum of the bounds on the light sheets and is therefore of order the entropy of the black hole, $S.$

It is also noteworthy that a causal patch is the largest region of the ERB that has a covariant entropy bound.

Now I will define a global entropy of the ERB---call it $S_{ERB}$---by the sum of the entropies of all the non-overlapping causal patches that fill the ERB. We can quibble later about calling it an entropy; for now it's just a definition. Recall that the total coordinate length of the ERB at time $t$ is $t.$ Therefore number of non-overlapping  causal patches that fit into the ERB is

\be
N_{cp} = t/\Delta t
\ee
with $\Delta t$ given by \ref{coordD}. Thus

\be
N_{cp}= (D-1 )\frac{t}{l}\sqrt{|f(r_f)|}
\ee
and the global entropy is,

\be
S_{ERB} = \frac{A}{G}N_{cp} = \dfrac{(D-1)V}{lG}.
\label{serbb}
\ee

The meaning of $S_{ERB} $ becomes clear when we compare \ref{serbb} with \ref{c-v relation}. The global entropy precisely matches the complexity of the state at time $t$ including  the factor o $(D-1).$

\be
S_{ERB} = \c
\label{serb = c}
\ee

The same thing happens for the  Schwarzschild black hole including the  matching of the factor $(D-3)$  in
\ref{c-v relation-sch-simple}.

Counting causal patches weighted by their entropy gives a new way to think about complexity, which is more general than identifying it with the volume of the ERB. Equation  \ref{serb = c} is correct for ADS black holes of any dimensionality, mass, and ADS radius. It is also true for all Schwarzschild black holes in any dimension.

However, what I have derived is not a complete demonstration of the universality of
\ref{serb = c} because it does not address charged or rotating black holes\footnote{I am grateful to Don Marolf for a discussion of these issues.}. It is not hard to see that the existence of a maximal slice exists for these cases, and that the growth of the volume of the ERB is proportional to $TS.$ What I am unable to verify is the connection between complexity and causal patches which depends on the location of the singularity. Classically the singularities of charged and rotating black holes are located beyond the inner horizons which are known to be unstable to both classical and quantum fluctuations. It is expected that the real singularity is space-like but I don't know enough about it to carry out the analysis of causal patches.

All of these considerations  of this part of the lecture were derived assuming the complexity increases linearly with time. However this cannot go on forever
since the complexity is bounded by

\be
\c_{max} \sim e^S.
\ee
This implies a limit on the growth the an ERB, which from a classical viewpoint is completely surprising.
However, classically $S$ is infinite (there is an $\hbar$ in the denominator of the Bekenstein Hawking formula), implying that an ERB can be arbitrarily long. Quantum mechanically the length is  bounded; in Planck units it cannot grow longer than the exponential of its area. In some way non-perturbative quantum fluctuations  build up and cause the collapse of the ERB. We can gain some more insight when we review the facts about complexity.

\bn

\sc
\section{Quantum Circuits and Tensor Networks}

\subsection{Cable Networks}

I've used the language of qubits without ever  specifying what the qubits are. I'll give an intuitive discussion of a string theory example. There is no pretense to rigor or precision; my purpose is to lay out a qualitative picture of how  qubits can emerge from string theory and how they behave.
 The example is the  black hole in $ADS(3)\times S(3) \times T(4), $ for which the microscopic degrees of freedom are known. They are a system of D5 and D1-branes. Their numbers are $Q_5$ and $Q_1.$ The D5-branes wrap the angular direction of ADS ($\theta$) and fill the 4-torus. The D1-branes also wrap $\theta$ but are localized on the torus. Their locations define a set of four scalar field multiplets $\phi(\theta).$ Each multiplet has multiplicity $Q_1Q_5.$ We will denote the multiplicity by $K.$

According to the long string picture of \cite{Maldacena:1996ds}
the boundary conditions are

\bea
\phi_n(2\pi ) \eq \phi_{n+1}(0)  \cr  \cr
\phi_{K+1} \eq  \phi_1.
\eea

\begin{figure}[h!]
\begin{center}
\includegraphics[scale=.3]{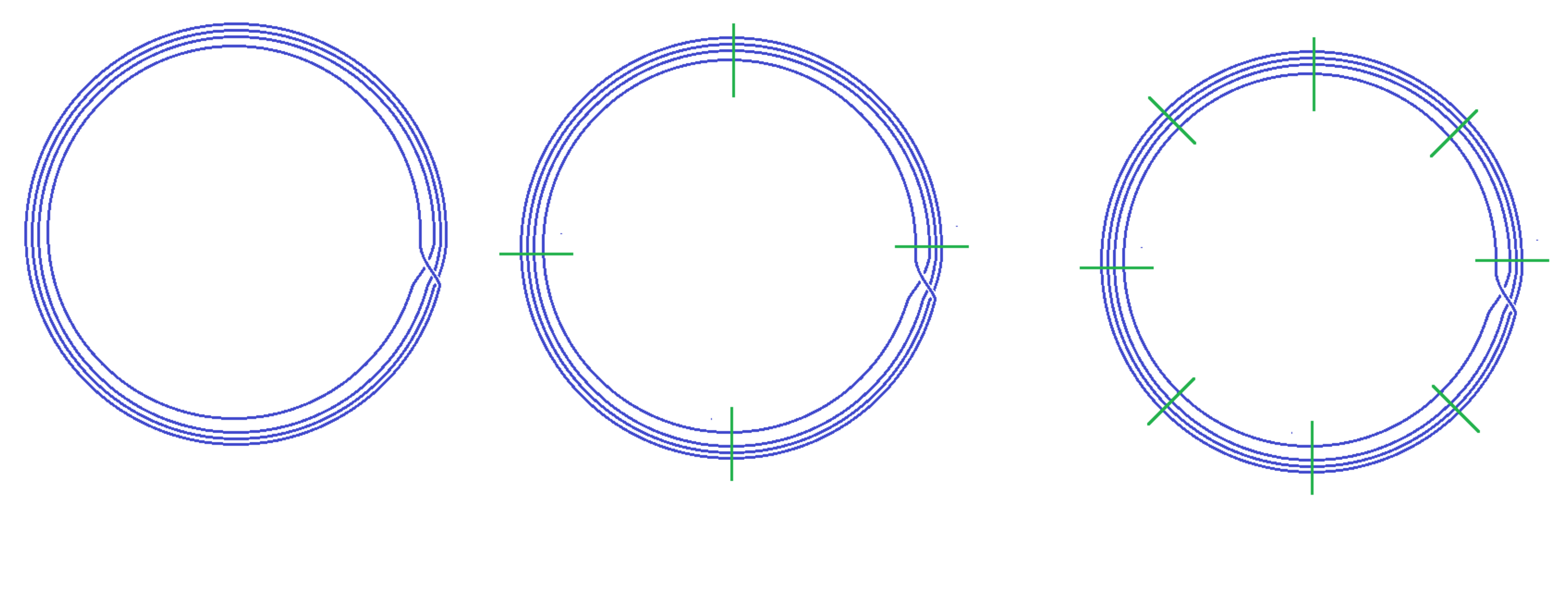}
\caption{The long string model of the D1-D5 system provides a model for black hole qubits. In the left panel the black hole is at the Hawking-Page transition. The thermal wavelength in the angular coordinate is $2\pi$ and each coil has of order one bit of entropy.  When the temperature is increased the thermal wavelength decreases. Fractional winds of the string become the qubits. The qubit-like segments only interact directly if they are in the same or neighboring angular intervals. }
\label{K}
\end{center}
\end{figure}

Effectively the fields describe a single string of length $K.$ But one should keep in mind that it wraps the $\theta$ direction $K$ times. In figure \ref{K} the configuration is sketched.

Now let us consider a ``unit black hole" at the Hawking-Page transition, or just above it. The thermal wavelength on the string corresponds to a single $2\pi$ wind of the string so that each coil carries of order a single bit of entropy. Evidently each coil can be identified with a few qubits.
Since there are $K$ coils the total entropy is of order $K.$
The qubits can be identified with the string mode-operators for excitations of thermal wavelength. $Z$ is the occupation number which we can truncate at $0$ and $1,$  $X \pm iY$ create and annihilate a quantum. Since there are several scalar fields there are several qubits per coil.

Because each wind of the long string passes through the same angular location, they can be expected to interact. This means that there is no locality in the effective Hamiltonian; every qubit interacts with every other qubit. I will not try to derive the form of the effective Hamiltonian here. The assumption is that each wind constitutes a qubit or a small number of qubits, and that the interaction Hamiltonian is built from interactions of pairs of qubits, with a term for each pair.

Next consider raising the temperature and the entropy. The thermal wavelength on the string becomes shorter, say $2\pi/n$ and the entropy increases by a factor of $n.$ The qubits are no longer full $2\pi$ winds of the long string, but instead occupy angular intervals of length $2\pi/n.$   By ordinary locality qubits at non-adjacent angular intervals don't interact. Therefore the Hamiltonian has a degree of angular locality, but all qubits within a angular interval interact with one another.

\begin{figure}[h!]
\begin{center}
\includegraphics[scale=.3]{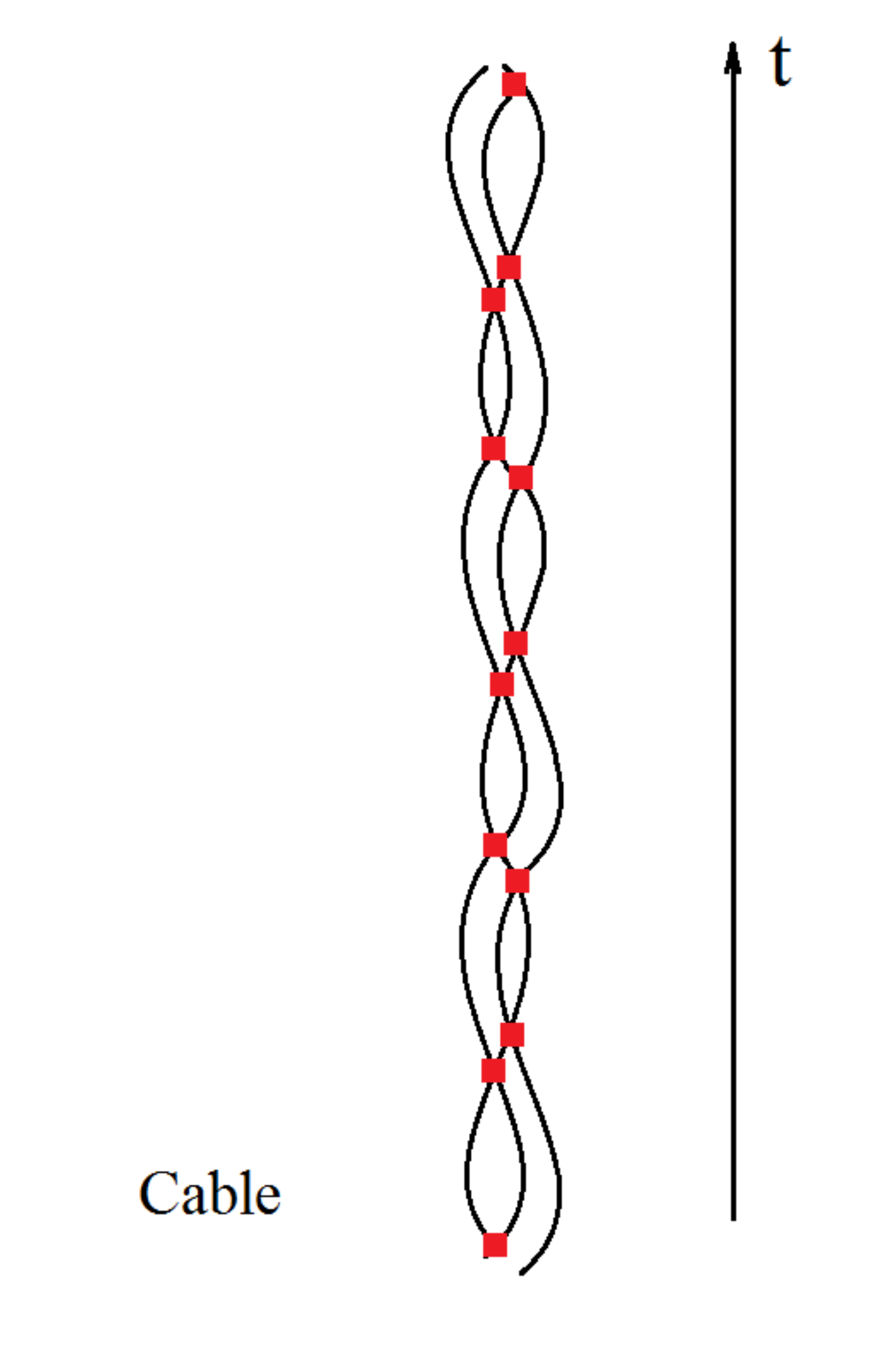}
\caption{A quantum circuit of one cable. All qubits within the cable can interact via two-qubit gates. }
\label{L}
\end{center}
\end{figure}

This picture suggests a structure for the quantum circuit describing the black hole evolution. Let's start with a ``unit" black hole  at the Hawking Page transition. The quantum circuits for this case should have no sense of spatial locality but should allow interaction any pair of qubits---in other words they are $2$-local. The quantum circuit for the evolution would have a structure shown schematically in figure \ref{L}.

I'll call a circuit of this kind a cable for obvious reasons.

\begin{figure}[h!]
\begin{center}
\includegraphics[scale=.3]{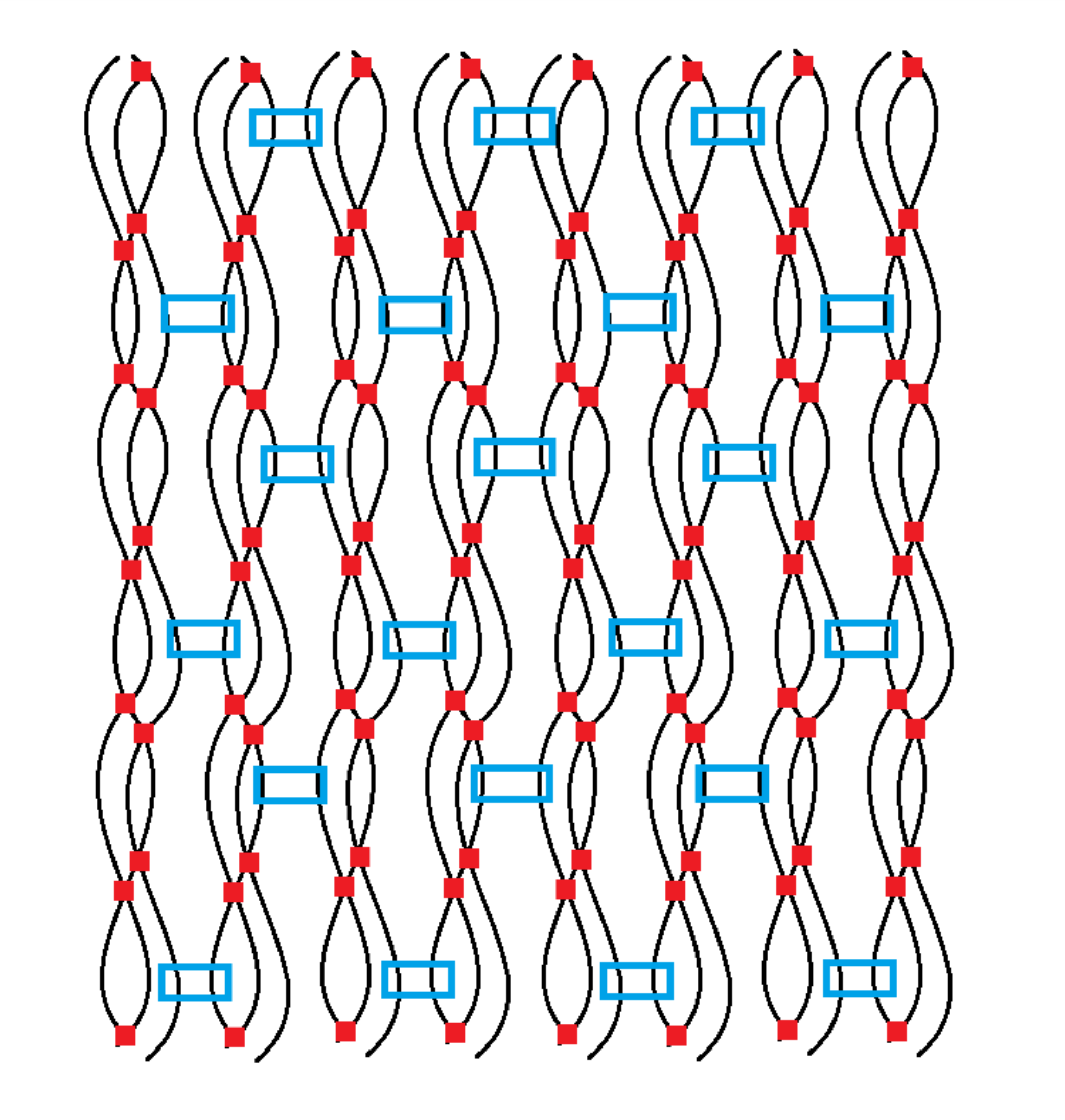}
\caption{ When the temperature is raised multiple cables combine to form a cable network.}
\label{M}
\end{center}
\end{figure}

When the temperature is raised the number of qubits increases. Groups of them associated with the same angular interval group together into cables, but the cables at different angular positions organize themselves into a local lattice structure. This is  shown in figure \ref{M}.

Within a cable there is no manifest sense of locality, but the interactions among the cables are organized by nearest neighbor interactions. I will call such a quantum circuit a ``cable network."

The evolution  of a quantum circuit begins with an initial state of the qubits---usually a fairly simple state. In the application to one-sided black holes the initial state represents the infalling matter. The evolution proceeds from there, becoming increasingly complex. Propagating the circuit for a time $t$ gives a representation of the state at time $t.$

\begin{figure}[h!]
\begin{center}
\includegraphics[scale=.5]{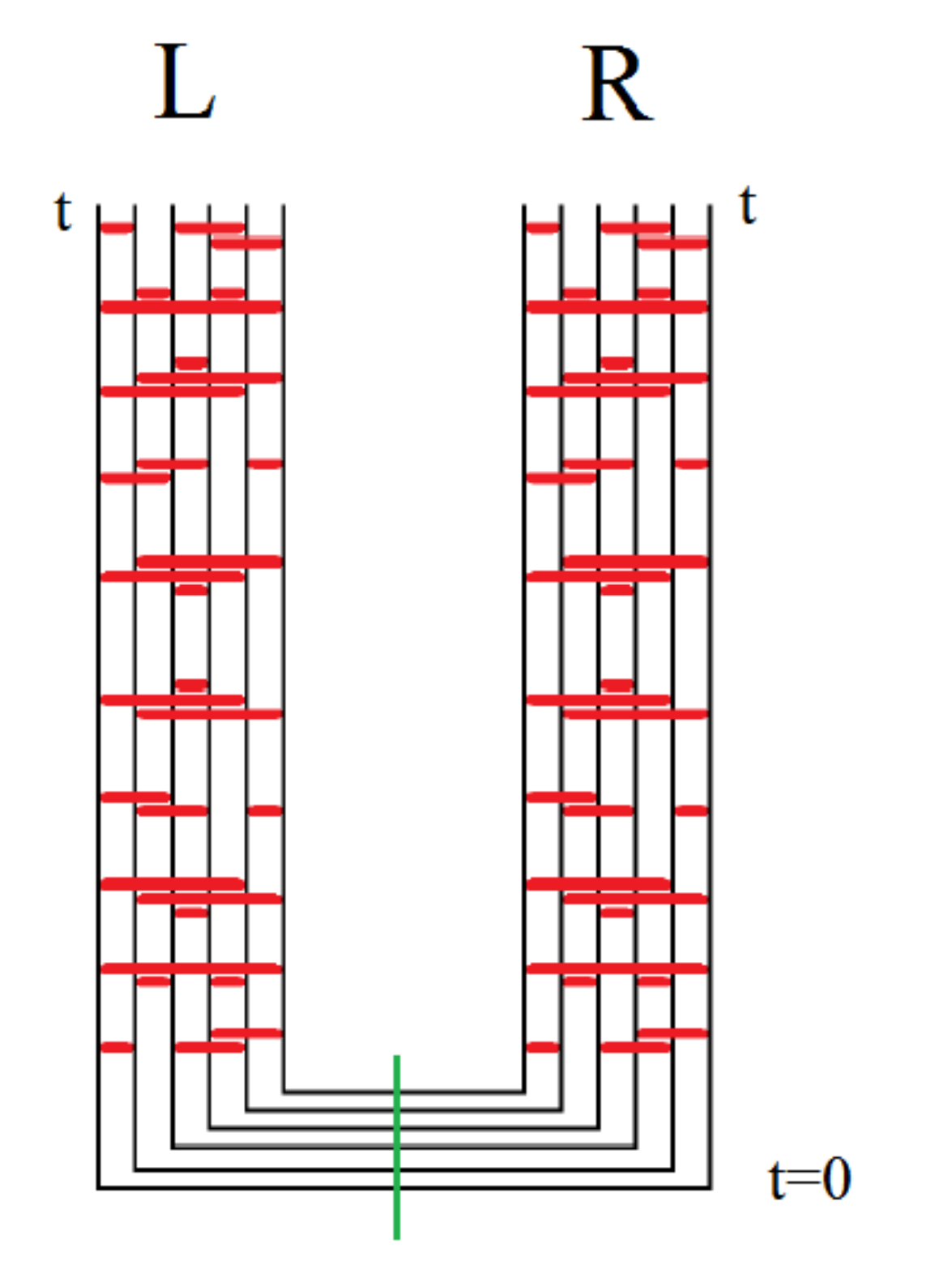}
\caption{ Quantum circuit modeling the evolution of the TFD state. The initial state is a product of Bell pairs shared between the left
and right sides. Each side evolves independently of the other.}
\label{N}
\end{center}
\end{figure}

Let us consider the QCs representing the two-sided system. Obviously they consist of two non-interacting quantum circuits of the kind described above. For example the case of two unit black holes (at the Hawking Page point) consists of two noninteracting cables. The initial state of the circuit should model the TFD state. In a simple description we can assume the qubits on the left and right side a paired into maximally entangled Bell pairs. The quantum circuit has the U-shaped form shown in \ref{N}.

The vertical sides can propagate for a time $t$ to represent the evolution of the TFD up to time $t.$

\subsection{Quantum Circuits are Tensor Networks}

The difference between quantum circuits and tensor networks is one of perspective. Quantum circuits provide models for the time evolution of systems. Of course we can be more abstract and just say that they give a method of generating quantum states. But I will think of them as models for evolution. The width of the circuit is the number of qubits and the depth is the amount of time the system has evolved.

Tensor networks are thought  of as existing in bulk space--at a fixed instant which represent the state at that  time \cite{vidal}\cite{Swingle:2012wq}\cite{Hartman:2013qma}. If one wants to vary the time one updates the tensor network by hand.

\begin{figure}[h!]
\begin{center}
\includegraphics[scale=.5]{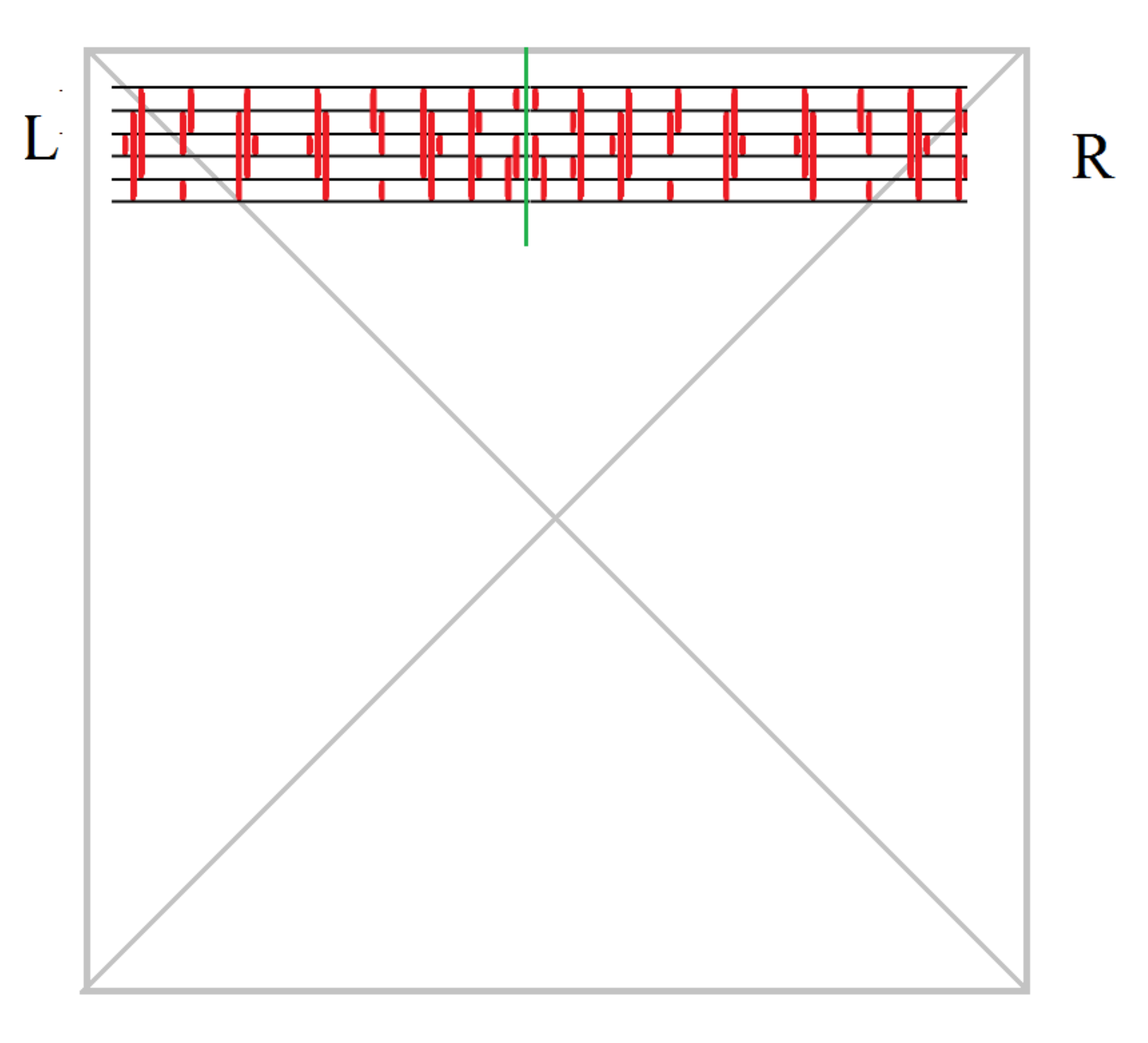}
\caption{The quantum circuit in figure \ref{N} can be lifted up to the maximal slice where it defines a tensor network
description of the ERB. }
\label{O}
\end{center}
\end{figure}

But from an abstract point of view tensor networks and quantum circuits are close relatives. In fact a quantum circuit is a particular type of tensor network in which the vertices are unitary matrices. Indeed there is evidence that the TN representation of many states is just the QC that one would use to evolve the state from a simple starting point. In the context of black holes it seems that the TN representing the state of an ERB is exactly the QC representing the evolution up to that time. We can take the example of the QC in figure \ref{N}. All we have to do to make a TN description of the ERB at time $t$ is to lift  it so that it is stretched across the space between the two endpoints. A crude sketch of this is in figure \ref{O}.

More precisely the TN can be laid out to fill the maximal surface connecting the boundary points  at $t$ on the left and right side. In this way we represent the state at $t$ by a tensor network in the manner of Hartman and Maldacena. As time increases we  add more to the ends of the TN in the left and right corners.  The minimal quantum circuit will translate in this way to the minimum tensor network representation of the state.

We may think of the TNs as networks of cables, each containing $K$ strands. The cable networks
 have a coarse-grained geometry to them. Roughly speaking the metric represents the minimum number of edges connecting nearby points. It may be thought of as a discretization of the spatial geometry of the ERB. However this only makes sense for a black hole far above the Hawking  page transition. The geometry is coarse-grained so that the cable thickness represent a resolution no finer than the ADS scale $l$. Finer resolution is encoded in the substructure of the cables in a non-local manner.

The geometry of the TN is connected with the Einstein geometry of the maximal slice. Roberts, Stanford, and I \cite{Roberts:2014isa}\cite{Stanford:2014jda} have devoted several papers to confirming the following proposition:

\bn

The geometry of the minimal tensor network (equivalently the minimal quantum circuit)
matches the Einstein geometry of the ERB in a coarse grained sense, with a spatial resolution of $l,$  or in the Schwarzschild case $r_H.$

 \section{ Nonlinearity}\label{S nonlinearity}

It's not obvious that   ER=EPR  is consistent with the principles of  quantum mechanics---a very rigid structure that does not happily tolerate change.  ER=EPR does seem to require nonlinear modifications of quantum mechanics behind the horizon\footnote{A number of authors have expressed the opinion that quantum mechanics may have to be modified in some nonlinear way behind the horizon. See for example \cite{Verlinde:2012cy}\cite{Marolf:2013dba}\cite{Papadodimas:2012aq}.}. I believe this is correct insofar as one is trying to describe the entire spacetime \it globally, \rm  but the non-linearities are not observable within a causal patch. An example of nonlinear features of a global description is cloning.

The original cloning question was first discussed long ago \cite{Susskind:1993mu}\cite{Hayden:2007cs}.  One can take two views:
The first is a ``tree-in-the-forest" viewpoint; If no one can see the cloning then it doesn't happen: The other is that cloning is a necessary feature of any global description that includes both the exterior and the interior of the black hole, but restrictions of causality never allow the cloning to be seen from a single causal patch. The second point of view may be the more fruitful one.

To illustrate the limits that the no-cloning theorem imposes in an ordinary context consider a
classical Turing machine. The machine is an ``updater" which takes a (classical) state and updates it to a next state according to some instruction programmed into the machine. A very simple example would involve a single classical bit that can be in either of two states, up or down. The instruction could be flip the bit:

\bea
(\uparrow)  &\to & (\downarrow) \cr \cr
(\downarrow) &\to & (\uparrow)
\label{cgate}
\eea
Thus the machine would produce an evolution (in time) that consists of a temporal sequence of configurations.  The history is completely determined by the initial  configuration which is either up or down.

\begin{figure}[h!]
\begin{center}
\includegraphics[scale=.3]{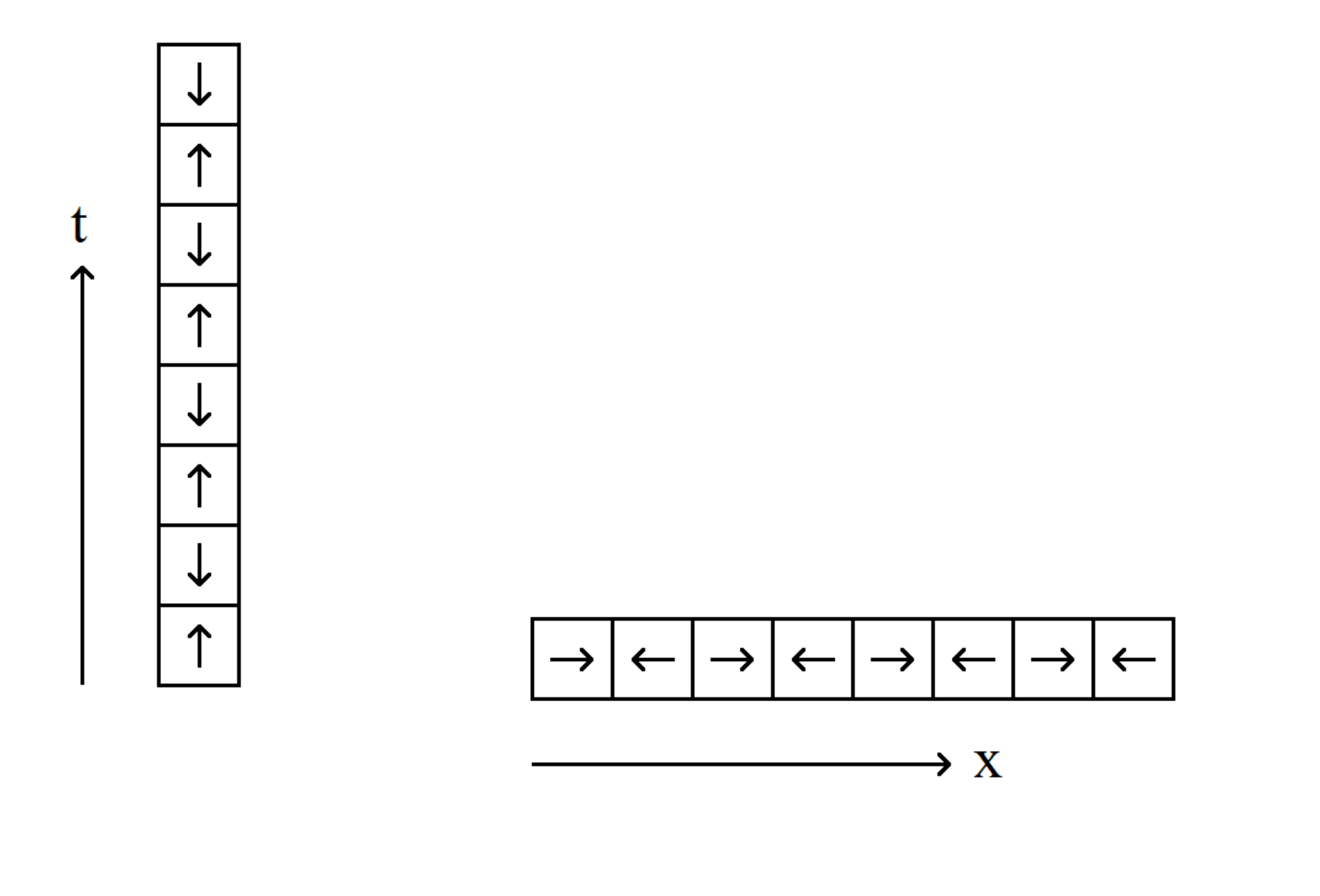}
\caption{A classical Turing machine implements a sequence of configurations of a c-bit. The machine produces a record of the computation in the form of a tape. The tape records the history of the single c-bit as a spatial sequence of c-bits.}
\label{turing}
\end{center}
\end{figure}

What is the output of the machine after some number of steps? One answer is that it is the final state, either an up or a down. But we can also suppose that the machine produces a
tape that records the history as a series of up and down arrows. Figure \ref{turing}  \ illustrates the point: the upper half of the figure shows the time sequence of states: the lower half shows the physical record laid out in space.

\begin{figure}[h!]
\begin{center}
\includegraphics[scale=.3]{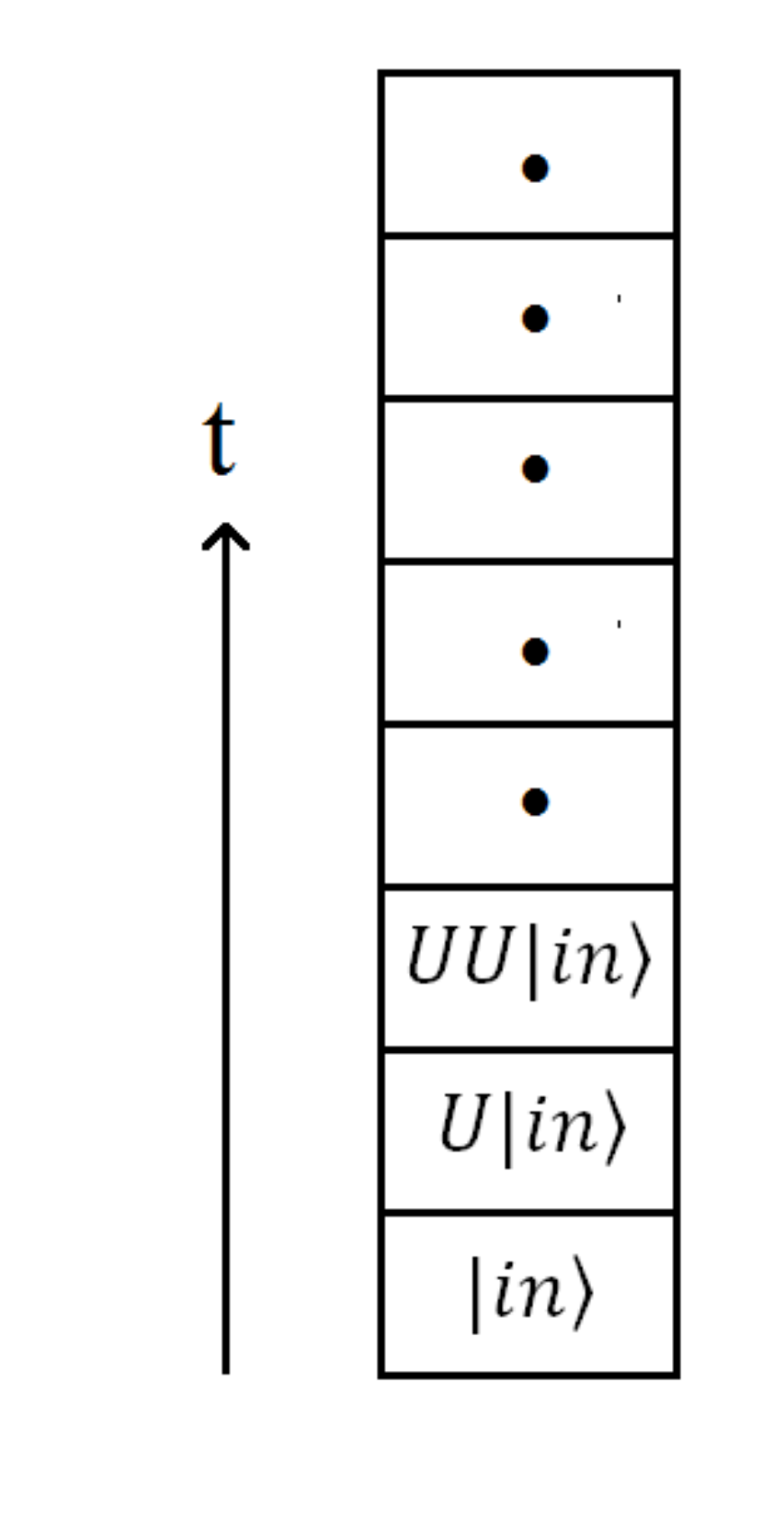}
\caption{A quantum Turing machine acts on an initial state and sequentially transforms it by unitary operations. It is not possible to record the sequence of states in the form of a spatial array of qubits. That would be non-linear cloning which is forbidden in quantum mechanics.}
\label{quring}
\end{center}
\end{figure}

Now consider the quantum version in which the classical  bit is replaced by a qubit and the instruction is a unitary one-qubit gate. As an example the gate could be the quantum analog of \ref{cgate},
\bea
|\uparrow\ra  &\to & |\downarrow\ra \cr \cr
|\downarrow\ra &\to & |\uparrow\ra
\eea
but now the initial state is a general superposition,

\be
|initial\ra = a|\uparrow\ra +b|\downarrow\ra.
\ee
Once again the machine generates a temporal sequence of (quantum) states indicated in figure \ref{quring}.
However, in the quantum case it is not possible for the machine to output a record of the sequence of states in the form of  a
 physical  sequence of qubits in space. Any such record would constitute a version of quantum cloning and would violate the linearity of quantum mechanics\footnote{One usually thinks of cloning as creating an identical copy of an original. In this case the copy is a unitary transform of the original. }.

 And yet that is exactly what is happening in figure \ref{O}. The history of the quantum circuit is recorded on the maximal slice.

Returning to the unspooling of an ERB, figures \ref{N} \ref{O} illustrate how
 as $t$ advances, more and more of the final slice is emerges within the ERB. From the figure one  sees that an increasing number of casual patches are created in the upper corner of the geometry. Let us consider the quantum circuit version of evolution. One starts with some  state in the Hilbert space of the  infalling matter. Call it $|s\ra.$ After one unit of time ($l$ for ADS, $r_H$ for Schwarzschild   ) the QC has updated the state,

  \be
  |s\ra \to G_1 |s\ra .
  \ee
   where $G_1$ is a product of about $S/2$ gates. That's straightforward, but in the tensor network version the ERB contains two causal patches, the first related to the state $|s\ra$ and the second to the state $G_1 |s\ra.$ Although the two causal patches may not be in the same state, they are unitarily related by $G_1.$

\begin{figure}[h!]
\begin{center}
\includegraphics[scale=.3]{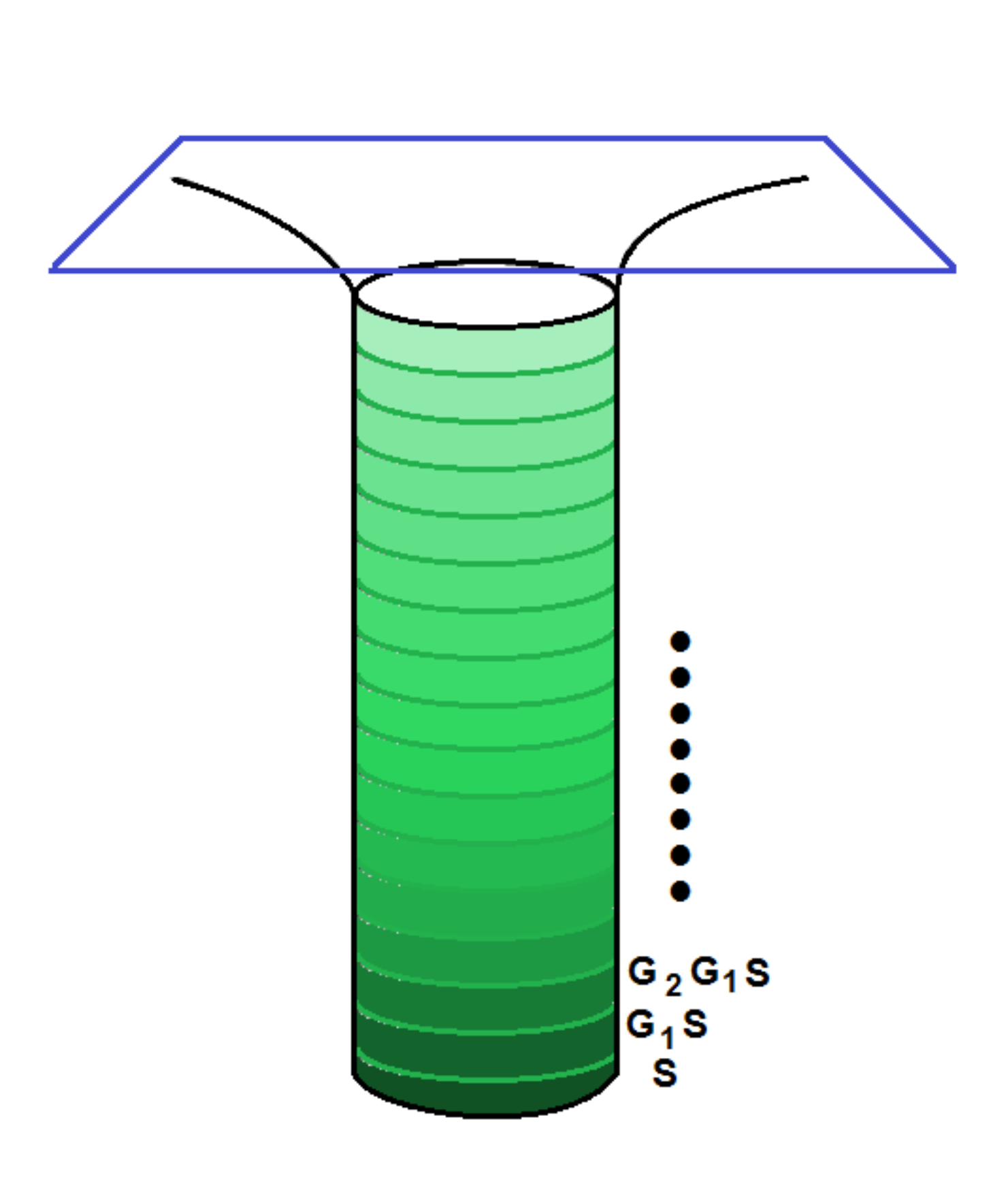}
\caption{As causal patches are unspooled in an ERB the initial state is repetitively cloned.}
\label{cloning}
\end{center}
\end{figure}

   As the state continues to evolve,

 \be
  |s\ra \to G_1 |s\ra \to G_2 G_1 |s\ra  \to G_3G_2 G_1 |s\ra  ...
  \ee
causal patches are unspooled in states that are all unitarily related the first patch. See figure \ref{cloning}
   Each causal patch in figure \ref{cloning} contains a unitarily equivalent replica of the initial state $|s\ra.$
In other words the history has been laid out spatially in exactly the manner that constitutes nonlinear cloning. Thus to summarize:

\bn

\bi
\item An Einstein-Rosen bridge is an array of causal patches. Each causal patch is a clone of the first patch which encodes the infalling state that created the black hole in the first place. If matter falls in later it is repetitively cloned afterwards.
The entire ERB may be thought of as a record of the history of the black hole laid out spatially along the ERB.
\ei

\bn

 It is also clear that the global state does not exist in the Hilbert space of the dual CFT. It exists as a \it history of CFT states.\rm  \ The complexity, which equals $S_{ERB},$ is not the entropy  of a state, but of the sequence of states reflected in the growing ERB.

An interesting point is that if observable cloning were possible, wave functions would be measurable, as would properties like entanglement and complexity. It follows that NP-hard problems could be solved easily. Unfortunately this is not the case.

 \subsection{Is ERB Volume Observable?}

Is
the volume of an ERB  an observable, i.e., is it an hermitian operator?  Here are two conflicting answers:

\bn

\bi
\item Yes. The maximum volume spatial slice is defined in a gauge invariant way. It's volume can be expressed in terms of the metric on the slice, and in canonical quantum gravity the metric is a hermitian operator. For example we can use the Wheeler DeWitt formalism  to define a wave function, and gauge-fix it to the maximal slice. The volume of the slice would be a hermitian operator acting on the WDW wave function.

\bn

\item  No. If the volume of the slice is encoded as the complexity of the quantum state then it is a property of the wave function, but not a hermitian operator. It is like entanglement in that respect.

    \ei

    To see that complexity is not an observable consider two states of the two-sided system. The first state has a shock wave coming from the left \cite{Shenker:2013pqa}\cite{Shenker:2013yza}. The shockwave was created by the precursor $W_L$ and has complexity $\c > S\log S.$ \ The other state is exactly the same except that the shockwave was created on the right side by the precursor $W_R.$ It also has complexity $\c.$ The two shockwave geometries are classically easily distinguishable; they have macroscopically different geometries.

     Quantum mechanics allows us to linearly superpose these states. According to the linearity principle any observable which has a definite value, the same in both states, must have the same definite value in the superposition.
However, in general  a superposition will have complexity very different from either of the states being superposed. In the case of the left and right shockwaves the superposition can only be made by a circuit much larger  than the circuits for either of the two component states implying that its complexity is much larger.  Therefore complexity cannot be represented by a linear hermitian operator.

\bn

The best way to decide if the volume of an ERB is an observable is to ask if it can be observed---in principle?
The answer is clearly  no.   Every observer is identified with a causal patch bounded by the observer's personal event horizon, i.e., the light sheets shown in figure \ref{I}.
  Larger  regions can never be observed. Therefore the length and volume of the ERB  is not observable. There is no reason for it to be mathematically described as a linear hermitian operator.
The non-observability of the global geometry is  related to the cloning issue in an obvious way.

\bn

\bf
Question: Should we expect the standard linear rules of quantum mechanics for observations in a single causal patch?

\bn

Answer: That's the big question. Let me rephrase it:
Black hole complementarity \cite{Susskind:1993if} was intended to apply to experiments carried out on scales much smaller than the black hole radius.
Consider Bob falling into an  ADS black hole carrying a small laboratory with him. The dimensions of the laboratory are much smaller than the ADS radius---in particular the area $a$ of the laboratory walls is much smaller than the area of the horizon. The question is: can Bob describe all experiments in his laboratory by a linear Hilbert space of dimension $\exp{s}$ where $s=a/4G?$  Questions like this are very difficult because they require spatial resolution on scales much smaller than the ADS radius.

There are obvious limits on the precision that Bob can achieve because of the finite amount of space and time in his causal patch. For example it would not be possible to resolve energy differences of order $e^{-s}$ unless the black hole is exponentially bigger than the lab.
The limitations I suspect come from large complexity. Let us ask whether all possible linear superpositions of the $e^s$ basis states of Bob's laboratory can fit within a single causal patch? The  area of the laboratory is much smaller than $l^2$ so the lab should fit transversely. To put it another way, the entropy in the laboratory is much less than the entropy of a single causal patch. At worst it is $s=a/4G$ if the laboratory is filled with a black hole.

\begin{figure}[h!]
\begin{center}
\includegraphics[scale=.3]{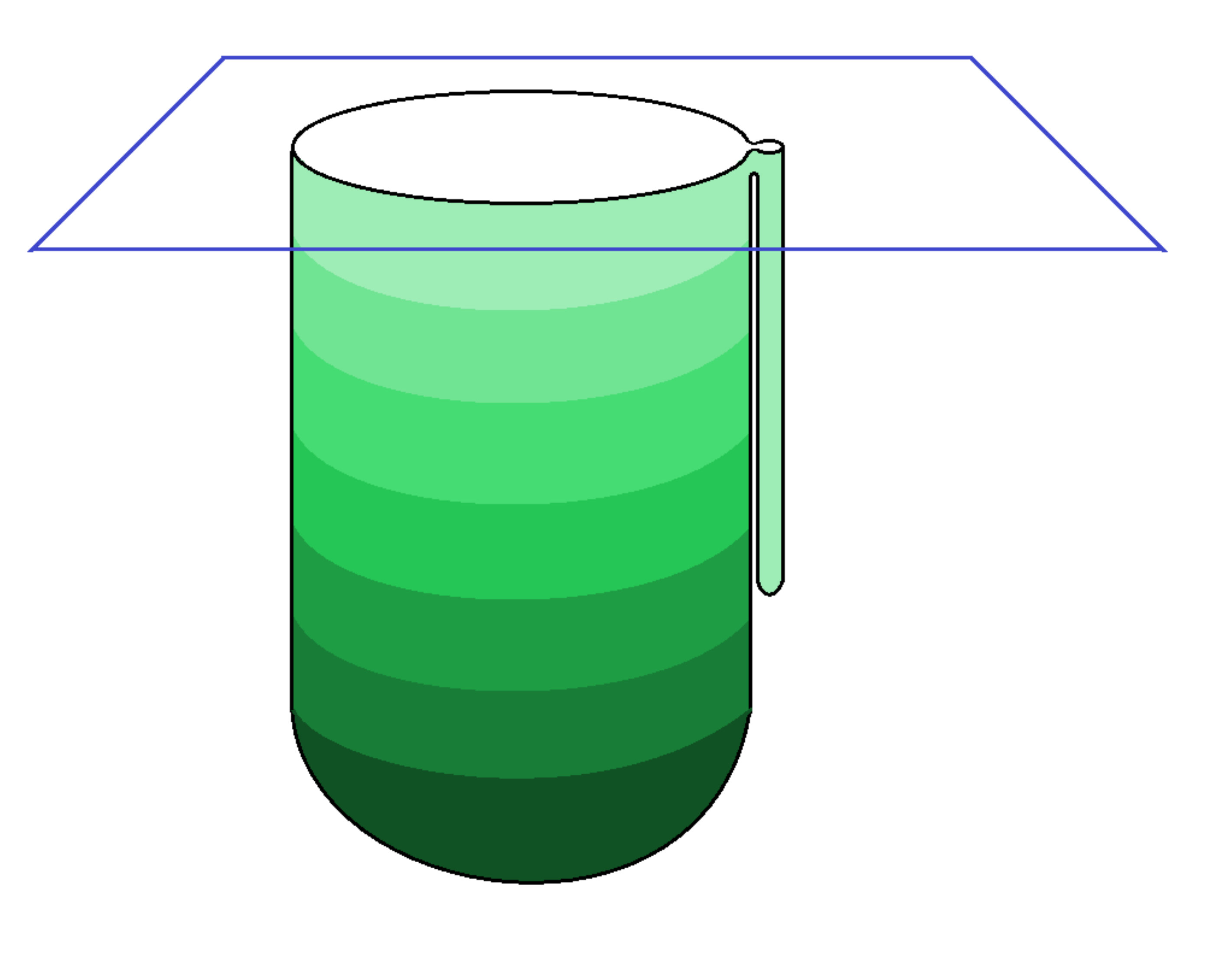}
\caption{A small old black hole merges with a large black hole. The small black hole has an area much smaller than the horizon area of the large black hole, but it may be too long to fit into a $CP(t).$}
\label{small-big}
\end{center}
\end{figure}

But  entropy does not tell the whole story. All but a tiny fraction of the Hilbert space is occupied by black holes of complexity $e^s$ which means that they have ERBs of exponentially large length. Let's suppose Bob carries in a small but very old black hole into a large young black hole.
 Figure \ref{small-big} shows Bob's small black hole merging with the original much bigger black hole.

The ERB of the small black hole will not fit into a causal patch of the bigger black hole unless $e^s<S.$ The problem is not any better if the original big black hole is old. We are asking if the small black hole can fit into the most recent causal patch that Bob himself falls into.
Thus it seems that if the entire Hilbert space of Bob's lab is to fit into a single causal patch, the size of the lab must satisfy the stringent requirement that

\be
a < 4G \log S
\label{limitation}
\ee

For a big galactic mass black hole the laboratory size  $a$ would have to be less than about $1,000$ Planck areas for all the states to fit into a causal patch.

In practice the limitation \ref{limitation} is irrelevant. Ordinary laboratory experiments do not involve black holes, and certainly not states with exponential complexity. It is possible that the tension just represents the limitations on confirming the rules of quantum mechanics in a limited space-time region. But it is conceptually interesting that every state of a basis may fit in a causal patch but the overwhelming majority of superpositions do not.

The thing that seems interesting to me is that the question of Bob's limitations may have a translation into a geometric question that could be answered by classical general relativity.

\rm

\subsection{The Geometry of Complexity }

In this part of the lecture the term \it geometry \rm  refers to the geometry of Hilbert space; not  to spacetime geometry.

There is a very beautiful way to think about complexity due to Michael Nielsen and collaborators \cite{Nielsen}\cite{Susskind:2014jwa}. It begins with the observation that the space of  (special) unitary operators on $K$ qubits is $\suk.$ The action of a sequence of gates  defines a trajectory in $\suk.$ The trajectory consists of a number of discrete intervals. The complexity of a unitary operator is the minimum number of such intervals connecting the identity operator with $u.$ In other words, with suitable definitions, the complexity is the length of the shortest trajectory of a certain kind, connecting the two points $I$ and $u$ on the manifold $\suk.$

What Nielsen did was to smooth things out  and define a continuous version of the quantum circuit, in a way that replaces discrete trajectories $\suk$ by smooth curves.  Moreover, a Riemannian metric can be defined so that  the minimal path of gates may be replaced by the minimal geodesic connecting points in $\suk.$

The sequence of gates is replaced by a Hamiltonian evolution,

\be
u = P e^{-i\int h(s)ds}
\ee
where $s$ is a path parameter and $P$ is the path-ordering instruction.

A Hamiltonian is called $k$-local if it is built as a sum of terms, each of which involves no more than $k$ qubits.
The geometry is designed so that path-segments built from $k$-local Hamiltonians  are short. But directions generated by Hamiltonians  containing products of greater than $k$ qubits pay a large cost in length. A cost parameter $q$ is introduced so that the metric along $k$-local directions  is not costly, but along other directions the metric is multiplied by $q.$

Nielsen argues that if $q>4^K$ the definition of complexity is insensitive to its precise value.

Let's assume that the actual Hamiltonian is  $k$-local. Then it is easy to prove that for short times the complexity of the time evolution operator increases linearly with time. The trajectory swept out by $e^{-iHt}$ is a geodesic if $H$ is  $k$-local. Moreover for some period of time it is the shortest geodesic. For how long that situation persists  is an important question. It cannot continue indefinitely because complexity has an upper bound. One way to see that is by counting parameters. Let's do it for the case of $k=2.$ A two-qubit gate is an element of $SU(4) $ and requires a specification of $15$ real numbers. A general unitary in $\suk$ requires $4^K-1$ real parameters. Therefore there are elements of $\suk$ which cannot be obtained by fewer than $(4^K-1)/15$ gates. On the other hand parameter counting suggests that  $(4^K-1)/15$ gates should be sufficient for any element of $\suk.$ It follows that the maximal complexity is of order $4^K.$

Now consider the shortest geodesic between $I$ and $u$ in complexity geometry. Begin by setting the cost parameter $q$ to unity. Then every point $u$ is a distance of order $1.$ Take the corresponding geodesic and evaluate its length in the geometry with general $q.$ It is obvious that it will not be bigger than order $q.$ Setting $q=4^K$ one sees that the shortest geodesic between any pair of point must be no bigger than $4^K.$

Now consider the geodesic swept out by $U(t) = e^{-iHt}$ where $H$ is the true Hamiltonian of the system. If one assumes that the spectrum of $H$ is chaotic (incommensurate levels) then the geodesic cannot be finite in total length. But since the maximum geodesic distance is $\sim 4^K$ it means that after a point there must be shorter geodesics connecting $U(t)$ and $I.$ The transition happens at a point on the cut-locus\footnote{see  Wikipedia---Cut locus.}. The geodesic $U(t) = e^{-iHt}$ intersects the cut locus at a distance no greater than $4^K.$

The reason that this is interesting is that reaching the cut point signals a breakdown of classical general relativity, as we will see.

The question of whether the properties of the region behind the horizon of ADS black holes are represented in the dual CFT has come up repeatedly. With the help of Nielsen's geometry I will
 give one more example of how non-linear the description of the interior can be when expressed in terms of the dual gauge theory. In defining complexity by Nielsen's geometric construction \cite{Nielsen} one runs into a peculiar situation. The complexity metric on the space of unitary operators $\suk$ is very distorted away from the conventional  metric\footnote{By the conventional metric I mean the metric invariant under left and right multiplication by unitary operators. }. In fact two unitary operators can be very close in the usual metric but extremely far in complexity. For example given a bounded Hermitian operator $h,$ and a small parameter $\epsilon <<1$ the operator $e^{-ih\epsilon} $ is very close to the identity in the usual metric on $\suk.$. But if $h$ is a hard operator, i.e., not $k$-local,
then the distance in the complexity metric may be of order $\epsilon q$ where $q$ is the penalty parameter. The penalty parameter in Nielsen's construction is order $4^K$ where $K$ is the number of qubits.

What is the implication of this large distance between operators that we ordinarily think of as close? It means that the minimal tensor networks for constructing $e^{-ih\epsilon}$ may be very large. One can say the same about states: the minimal tensor networks for the states $|\psi\ra$ and $|\phi\ra$ may be very different even if $$\la \phi|\psi\ra  $$ is very close to unity.

The implications are strange; a very different tensor network implies a very different interior geometry. Two gauge theory states can be very close in the usual sense---inner product close to one---but the corresponding global interiors may have radically different geometries.

An example would be the following. After the quantum recurrence time the TFD state will evolve to a state very close to itself. The recurrence may be to a state which is close in both the usual metric and the Nielsen metric. In that case the ERB geometry shortens back to the   TFD state with the horizons touching. But it will also return to states which are close in the usual metric, but very far in the Nielsen metric. In that case the ERB geometry will not be near the shortened TFD geometry, but will be very long.

Two states can be so close that

 \be
 \la \psi|\phi \ra = .9999999999.......
 \ee
 and the expectation values of all gauge theory observables will be almost indistinguishable, and yet the bulk ERB geometries  be extremely different.

  As far as I know this always involves exponentially long time scales and very large complexity.

\subsection{GHZ-branes}

Quantum mechanics is being geometrized in surprising new ways. Entanglement is area, complexity is volume, and more is surely in store. I want to discuss a possible new geometric structure related to measurement theory.
To set the context let's look at a simple exercise  involving qubits. Let there be two qubits called $A$ and $S$. The notation $A$ is short for apparatus, and maybe Alice. $S$ is a system which $A$ will measure.  The degrees of freedom are $\sigma_i$ for $A,$  and $s_i$ for $S.$

The initial state is an unentangled  product,
\bea
|\psi\ra \eq \left\{a|\downarrow\ra + b|\uparrow\ra \right\}_S \otimes |\downarrow\ra_A \cr \cr
&\equiv&  a|\downarrow\downarrow\ra + b|\uparrow\downarrow\ra
\label{state}
\eea
where the first entry represents the system $S$  and the second represents the apparatus $A$. Assume that when $A$ and $S$ interact the following unitary transformation takes place,

\bea
|\downarrow\downarrow\ra &\to& |\downarrow\downarrow\ra \cr \cr
|\uparrow\downarrow\ra &\to& |\uparrow\uparrow\ra \cr \cr
|\downarrow\uparrow\ra &\to& |\downarrow\uparrow\ra \cr \cr
|\uparrow\uparrow\ra &\to& |\uparrow\downarrow\ra
\label{transform}
\eea

After the interaction the state becomes,

\be
a|\downarrow\downarrow\ra + b|\uparrow\uparrow\ra.
\label{final state}
\ee
The result of the measurement is an entangled state. If the amplitudes $a$ and $b$ were approximately equal it is almost maximally entangled.

It is obvious that the thing which was measured is the $z$ component of the system-spin $s_z.$ One can read that off from the final density matrix of the system which is easily seen to be diagonal in the $s_z$ basis.

The final density matrix of the apparatus is also  diagonal in $\sigma_z$ basis. To the extent that we think of $A$ as an observer we can say that at the end of the experiment Alice finds herself in a state of definite $\sigma_z,$ but with classical probabilities $|a|^2$ and $|b|^2$ for the two values. What constitutes a measurement is the fact that Alice's state and the state of the system are completely correlated: if $s_z=\pm 1$ then $\sigma_z = \pm 1.$

According to the ER=EPR hypothesis the measurement has created a Einstein-Rosen bridge between the system and the apparatus. Of course for the simple system of a single qubit, the bridge has no meaning beyond the fact that $S$ and $A$ are entangled. No real Alice or Bob can jump into the qubits and check if there is an ERB.

Now lets suppose that $S$ is replaced by a large black hole in an initial pure state, and $A$ by an equally large system containing a  memory composed of enough qubits to register a complete set of observables of $S.$ We'll also introduce another observer, Bob, whose role is entirely passive at first. Bob knows the initial state and is capable of solving the Schrodinger equation, but does not interact with Alice or the black hole.

Alice will measure a complete set of observables of $S$ and record the results in her memory. The basis eigenstates of the black hole are labeled $|I\ra_S$ and the corresponding states of Alice's memory are $|I\ra_A.$

Let's first consider Bob's description of the $S,A$ interaction.
According to Bob the interaction leads to a
 highly entangled state with an ERB connecting $A$ and $S.$ For simplicity we can take it to be maximally entangled,

 \be
 |\Psi\ra = \sum_I |I\ra_S |I\ra_A
 \ee
 The ERB would be similar to an octopus, but Alice may choose to collapse herself into a black hole without losing any information.

Alice's view is different than Bob's. According to Alice, she has made a projective measurement and recorded the results in her memory. For her purposes the final state of  original black hole is a pure eigenstate of the measured operators. She also feels herself to be in a definite state. Therefore there cannot be an ERB connecting $A$ to $S.$ It seems that ER=EPR leads to a contradiction.

Can Alice confirm her prediction that there is no interior geometry? Suppose she tries to jump into the black hole carrying her memory with her. That's clearly not possible. The memory has at least as many degrees of freedom as the black hole. Entering the black hole with her memory would create an enormous back reaction and move the horizon far out from its original location.

Alice might choose the option of leaving her memory behind before she jumps in. Let's say she gives it to Bob and then jumps in. All that has happened is that she has changed places with Bob. When Alice jumps in she will be entering a black hole with an ERB connecting the black hole to Bob.

\bn

Now let's go on to the tripartite case.
Let's return to the situation where Alice has just made a complete measurement and is entangled with the black hole. Now Bob, who also has a big memory, enters the game and makes a complete measurement of Alice's memory. The operators he measures are exactly the ones in which Alice's density matrix is diagonal. First let's consider the case where the system, Alice, and Bob are all single qubits. The state just before Bob measures Alice is given by,

\be
 \left\{ a|\downarrow\downarrow\ra + b|\uparrow\uparrow\ra \right\} \otimes |\downarrow\ra_B
\ee

After Bob's measurement the state becomes,

\be
|ghz\ra = \left\{ a|\downarrow\downarrow\downarrow\ra + b|\uparrow\uparrow\uparrow\ra \right\}
\label{ghz}
\ee

This is the Greenberger, Horne, Zeilinger ( $ghz$ ) tripartite entangled state \cite{greenberger} of three qubits. It has some surprising properties. First of all each of the qubits is maximally entangled with the union of the other two. This is easy to see by calculating the one-qubit density matrices which are all maximally mixed and the
 entropy of each of the qubits is $\log{2}.$

But there is no entanglement between any two qubits. The two-qubit density matrices have the form,

\be
\rho = |\uparrow\uparrow\ra \la \uparrow\uparrow| +|\downarrow\downarrow\ra \la \downarrow\downarrow|
\label{separable}
\ee
 \ref{separable} which is separable, meaning that it is the sum of projection operators onto pure states. Such density matrices have zero entanglement. Thus  each qubit is maximally entangled with the union of the other two, but no two qubits are at all entangled.

Now let us consider the analogous state of three black holes. This is the state that would result after Bob measures Alice's memory and records the result in his own memory, which he then collapses to a third black hole. The analog of the $ghz$ state becomes

\be
|GHZ\ra = \sum_I |I, I, I\ra
\label{GHZ}
\ee
where the label $I$ runs over a complete set of black hole states in the measurement basis determined by Alice's choice of experiment.
If we assume that the black holes can each be described as a collection of $K$ qubits, then $|GHZ\ra$ can be written in a very simple form,

\be
|GHZ\ra = |ghz\ra^{\otimes K},
\ee
in other words as a tensor product of $K$  $ghz$-triplets. This is the analog of writing a bipartite state in terms of Bell pairs.

The $GHZ$ state of three black holes has the property that:

\bn

Each black hole is maximally entangled with the union of the other two, but there is no entanglement between any pair of black holes.

\bn

We can imagine creating a $GHZ$ triplet of black holes by starting with $K$  $ghz$ qubit-triplets. Each triplet is shared between Alice, Bob, and Charlie who separate  their shares and then collapse them to black holes.  To draw a tensor network I will use a basic representation of $ghz$ as three lines joined at a point. The $GHZ$ state would be represented by figure \ref{GHZ}.

\begin{figure}[h!]
\begin{center}
\includegraphics[scale=.3]{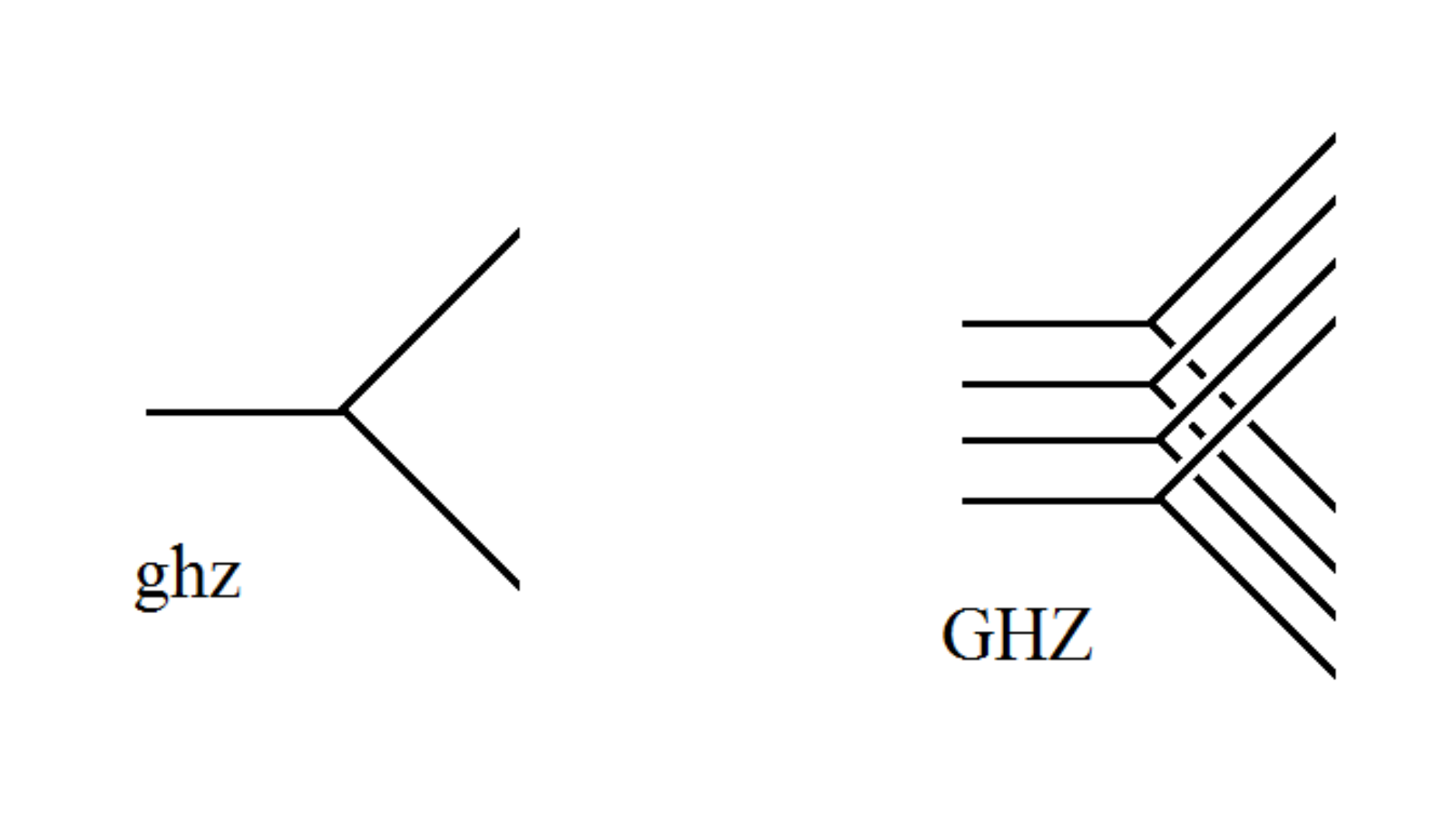}
\caption{ The left panel shows a basic $ghz$ triplet of qubits. On the right side many $ghz $ triplets combine to form
a $GHZ$ state.}
\label{GHZ}
\end{center}
\end{figure}

\begin{figure}[h!]
\begin{center}
\includegraphics[scale=.3]{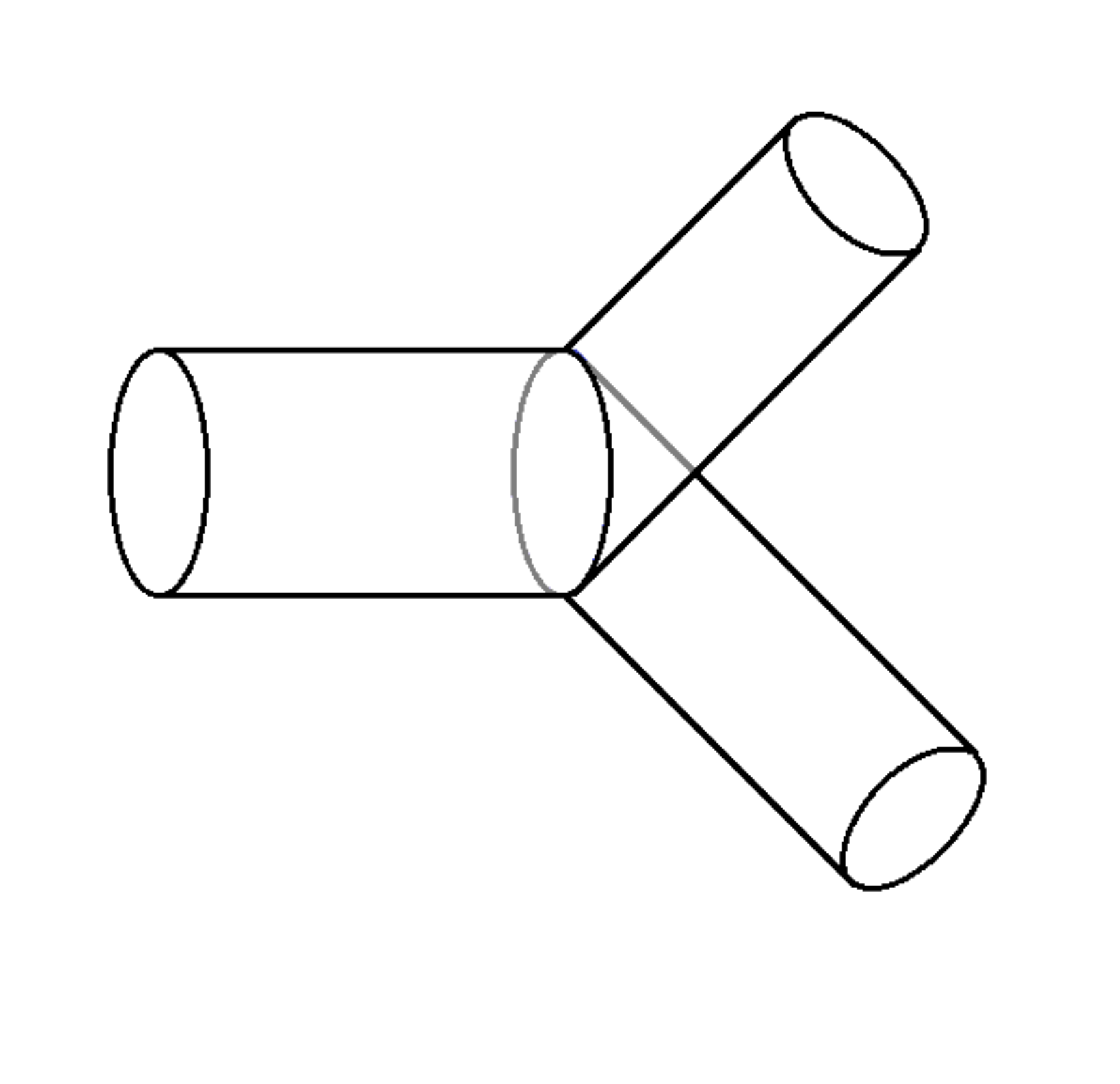}
\caption{The $GHZ$ state of three CFTs. Three black holes are joined in a macroscopic structure. }
\label{GHZ2}
\end{center}
\end{figure}

In the ADS context the state would involve three asymptotic boundaries joined together as in figure \ref{GHZ2}

Evidently the ERB connecting Alice, Bob, and Charlie is not a completely smooth geometry. There is a tripartite structure at the center of the ERB---call it a GHZ-brane---which
any  signal  passing  through the ERB must encounter\footnote{ One may guess that  smooth multi-black hole ERB state have no $GHZ$-type entanglement. The guess is supported by the calculations of Balasubramanian, Hayden, Maloney, Marolf and Ross,\cite{Balasubramanian:2014hda} where  the lack of $GHZ$ entanglement is explicitly shown for a class of smooth ERBs in $(2+1)$-dimensions.}.

The GHZ-brane must have some rather special properties.
For example, any signal originating from a local manipulation by one member of the triad cannot be detected by an observer falling through either of the other horizons. However if two of the members are brought together and manipulated, the signal may be detected by the third member as he or she falls through the horizon.

Obviously we don't have to stop with the tripartite case. If Dora comes along and measures Charlie a four-party entangled state results, and so on.
That is all I will say about GHZ-branes other than they are obviously worth study.

\section{Conclusion: Entanglement is not Enough}

What is it that makes quantum physics so different from classical physics? Two things: first is entanglement; the fact that you can know everything that can be known about a system and know nothing about its parts. The second distinguishing property of  quantum mechanics was pointed out by Feynman; namely the extraordinary potential complexity of quantum states.

Entanglement is not enough to understand the rich geometric structures that exist behind the horizon and which are predicted by general relativity. Entanglement entropy only  grows for a very short time, but the growth of Einstein-Rosen bridges is expected to last for a very long time. Encoding that growth in the quantum state requires quantum complexity.

One interesting question that this raises is exactly what does happen to a one-sided ADS black hole after  an extremely long time has elapsed. Eventually, after a doubly exponential quantum recurrence time, the black hole will return to its original state. The ERB will contract and debris will backflow toward the horizon.

  The behavior after a singly exponential time is more confusing. The complexity will become maximal, and the quantum state  generic.  The classical picture of eternal growth must break down but what replaces it is not obvious, perhaps firewalls. However real black holes will have evaporated long before this time.
\bn

\section*{Acknowledgements}

I am grateful to    Patrick Hayden, Don Marolf, Joe Polchinski,   Xiaoliang Qi,  Dan Roberts, Steve Shenker, Douglas Stanford,  and Ying Zhao, for numerous helpful and stimulating discussions.

This work was supported in
part by National Science Foundation grant 0756174 and by a grant from the John Templeton Foundation.
The opinions expressed in this publication are those of the author and do not necessarily
reflect the views of the John Templeton Foundation.

\end{document}